\RequirePackage{pdf14}
\documentclass[letterpaper]{article}
\usepackage[table]{xcolor}
\usepackage{array,setspace,mathrsfs,amsfonts,amsmath,yfonts,dsfont,bbm,colonequals,amscd}
\usepackage{./sections/jheppub}

\numberwithin{equation}{section}

\def\bea{\begin{eqnarray}}
\def\eea{\end{eqnarray}}

\def\PE{\mathrm{PE}}

\def\half{\frac12}
\def\S{{\mathbb S}}

\def \Rf{\mf{R}}

\def \beg#1{\begin{#1}} 
\def \be{\beg{equation}}
\def \bea{\beg{eqnarray}}
\def \eea{\end{eqnarray}}
\def \ee{\end{equation}}

\def \a{\alpha}
\def \b{\beta}

\def \PE{\mathrm{P.E.}}

\def \goodchi{\protect\raisebox{1pt}{$\chi$}}
\def \qq{\mathbbmtt{Q}\,}

\def \Lh{\widehat{L}}

\def \gf{\mf{g}}
\def \uf{\mf{u}}
\def \slf{\mf{sl}}
\def \suf{\mf{su}}

\def \hhf{\mf{h}}
\def \ceq{\colonequals}

\def \restr#1#2{{\left.\kern-\nulldelimiterspace#1\vphantom{\big|}\right|_{#2}}}

\def \hf{\frac12}

\def \mf{\mathfrak}

\def \wt{\widetilde}
\def \wh{\widehat}
\def \ol{\overline}

\def \Tr{{\rm Tr}}

\def \zb{{\bar z}}

\def \AA{\mathcal{A}}
\def \BB{\mathcal{B}}
\def \CC{\mathcal{C}}
\def \DD{\mathcal{D}}
\def \EE{\mathcal{E}}

\def \II{\mathcal{I}}
\def \JJ{\mathcal{J}}

\def \NN{\mathcal{N}}
\def \OO{\mathcal{O}}

\def \QQ{\mathcal{Q}}
\def \RR{\mathcal{R}}
\def \SS{\mathcal{S}}

\title{The superconformal index of theories of class \texorpdfstring{$\SS$}{S}}

\author[1]{Leonardo Rastelli}
\author[2,3,4]{and Shlomo S. Razamat}

\affiliation[1]{C.~N.~Yang Institute for Theoretical Physics, Stony Brook University, Stony Brook, NY 11794-3840, USA}
\affiliation[2]{Institute for Advanced Study, Einstein Dr., Princeton, NJ 08540, USA}
\affiliation[3]{NHETC,Rutgers University, Piscataway, NJ 08854, USA}
\affiliation[4]{Department of Physics, Technion, Haifa, 32000, Israel}

\bigskip
\abstract{  
We review different aspects of the superconformal index of  ${\cal N}=2$ superconformal theories of class ${\cal S}$. 
In particular we discuss the relation of the index of class ${\cal S}$ theories to topological QFTs and integrable models,
and review how this relation can be harnessed to completely determine the index. 

\

\noindent This is part of a combined review on  $2d$-$4d$ relations, edited by J.~Teschner. 

}


\begin{document}

\setcounter{tocdepth}{2}
\maketitle

\section{Introduction}

This volume surveys the $4d/2d$ relations that arise in the study
of class~${\cal S}$, the set of four-dimensional ${\cal N}=2$ supersymmetric field theories
 obtained by compactification of a six-dimensional $(2, 0)$ theory on a  punctured Riemann surface ${\cal C}$.\footnote{
See [V:1] in this volume for a general introduction to  class ${\cal S}$.}
There is an extensive $4d/2d$ dictionary relating  several protected observables of the four-dimensional theory ${T} [ {\cal C}]$
 to observables of  certain natural  theories defined on the associated surface ${\cal C}$.
 In this chapter we focus on the {\it superconformal index} of ${T} [ {\cal C}]$
 and on its  re-interpretation as a topological quantum field theory (TQFT) living on ${\cal C}$.
 We shall restrict our  discussion to  the subset of theories that enjoy conformal invariance, for which the
 general index  is well-defined.

 \smallskip
 
 The superconformal index, or index for short,  encodes some detailed information about the protected spectrum of a superconformal field theory. By construction, it is 
  invariant under exactly marginal deformations of the SCFT.
A basic item of the $4d/2d$ dictionary equates the conformal manifold of   ${T} [ {\cal C}]$ ({\it i.e.}, the space of its exactly marginal gauge couplings)
with the complex structure moduli of ${\cal C}$. We should  then expect on general grounds that the  index
is computed by a TQFT living on ${\cal C}$. A concrete description of this TQFT as an explicit
$2d$ theory is only available for certain specializations of the general index, in particular
the so-called Schur index corresponds to $q$-deformed two-dimensional Yang-Mills theory in the zero-area limit.
The TQFT viewpoint is however very 
 fruitful also in the general case. As different TQFT correlators
  compute the indices of different $4d$ theories,
 we are led to study consistency conditions in {\it theory space}. 
 This turns out to be  a very  effective strategy, which allows for
the complete determination of the general index for theories of class ${\cal S}$.

\section{The superconformal index}

Let us introduce the main character of this review. 
To a superconformal field theory in $d$ space-time dimensions one can associate its superconformal index~\cite{Romelsberger:2005eg,Kinney:2005ej},
which is nothing but the Witten index of the theory in radial quantization,
refined to keep track of a maximal set of commuting conserved quantum numbers
 $\{ C_i \}$,
\begin{equation}\label{genericind}
{\cal I} ( \mu_i)=\mathrm{Tr} \, (-1)^F
\prod_i \mu_i^{C_i} \, e^{-\beta \delta}
\,  , \quad \delta \colonequals \{ {\cal Q} \, , {\cal Q}^\dagger \} \,.
\end{equation}
The trace is taken over the  Hilbert space of the radially quantized theory on $\mathbb{S}^{d-1}$,
 $F$ is the fermion number and ${\cal Q}$ a chosen Poincar\'e supercharge.
In a given theory, the index is thus a function of the ``fugacities'' $\{ \mu_i \}$ that couple to the conserved charges $\{ C_i \}$.
The conserved charges are chosen as to commute with each other, with the chosen supercharge  ${\cal Q}$
and with its  conjugate (conformal) supercharge ${\cal Q}^\dagger$. If the theory is unitary, which we shall always assume,
 then $\delta \colonequals \{ {\cal Q} \, ,{\cal Q}^\dagger \} \geqslant 0$. 
 By a  familiar argument,
 the index counts (with signs) cohomology classes of ${\cal Q}$. 
Indeed the Hilbert space decomposes into  the subspace of states with $\delta =0$, which are automatically
killed by both ${\cal Q}$ and ${\cal Q}^\dagger$ (these are the ``harmonic representatives'' of the cohomology classes),
and  the subspace with $\delta \neq 0$, 
 where one can choose a basis such that all states belong
 to a pair  $(\psi,  {\cal Q} \psi )$,
 with ${\cal Q}^\dagger \psi = 0$. The paired states have the same charges $\{ C_i \}$ but opposite  statistics, 
 so their combined contribution to the trace vanishes. Since the trace in (\ref{genericind}) receives contributions
 only from the harmonic representatives,  the index is in fact independent of $\beta$. 
 The states with $\delta = 0$ are annihilated by some of the supercharges and as such
they belong to shortened representation of the superconformal algebra.

As the energy (conformal dimension) of a generic long multiplet of the superconformal algebra is lowered to  the unitarity bound, the long multiplet 
breaks up into  a direct sum of short multiplets,
containing states with $\delta=0$, but by continuity their total contribution to the index is zero.
So even  within the $\delta = 0 $ subspace there may be fermion/boson cancellations between states with the same charges $\{ C_i \}$,
associated to recombinations of short multiplets into long ones.
In fact one can equivalently characterize the index 
 as the most general invariant that counts  short multiplets,
up to the equivalence relation  setting to zero  combinations of short multiplets that have the right quantum numbers
to recombine into long ones \cite{Kinney:2005ej}. It follows, at least formally, that the index is invariant under changes of continuous
parameters of the theory  preserving superconformal invariance, {\it i.e.} it is constant
over the  {\it conformal manifold} of the theory. As the exactly marginal couplings are varied, long multiplets
may split into short ones or short multiplets recombine into long ones, but this is immaterial for the index. In other contexts,
the formal independence of supersymmetric indices on continuous parameters 
is known to fail, leading to rich wall-crossing phenomena.
 In our case, however,  we are dealing with theories  that have  a discrete spectrum of states,  and such that the  subspaces 
 with fixed values of the quantum numbers $\{ C_i \}$ are {\it finite-dimensional}, so the formal argument
is completely rigorous. The index is thus truly invariant under exactly marginal deformations preserving the full superconformal algebra of the model.

The superconformal index can  be defined for theories in various spacetime dimensions, and
with different amounts of superconformal symmetry. We have given the ``Hamiltonian'' definition  in terms
of a trace formula, but the index has an equivalent ``Lagrangian'' 
interpretation as a supersymmetric partition function on $\mathbb{S}^{d-1} \times \mathbb{S}^1$,
with twisted boundary conditions around the  ``temporal'' $\mathbb{S}^1$ to incorporate
the dependence on the various fugacities. See [V:5] in this volume for more details on this approach. 
Viewed as a partition function,
the index makes sense for  non-conformal theories, though in those cases it should be more
properly referred to as a {\it supersymmetric} index. One can show that such a partition function
is independent on the RG scale, so that the superconformal index
of a theory realized as the IR fixed point of  
some RG flow can be often computed using the non-conformal UV starting point 
of the flow~\cite{Romelsberger:2005eg,Romelsberger:2007ec,Festuccia:2011ws}. Examples  where this is a very useful strategy
include ${\cal N}=1$ gauge theories in four dimensions, and susy gauge theories
in three and two dimensions.  
The partition function interpretation is also useful to obtain the index in the presence of various BPS defects, by the techniques of supersymmetric localization.
In this review we will mostly stick to the trace interpretation of the index, and localization will not play a role.  
We will determine the index of  the ${\cal N}=2$ SCFTs of class ${\cal S}$  (even in the presence of certain BPS defects)  by a more abstract algebraic viewpoint.
A direct localization approach would not be an option since these theories do not generally admit a known Lagrangian description.

\

 We now specialize to the case of interest, namely ${\cal N}=2$ superconformal theories in four dimensions.
 The  ${\cal N}=2$  superconformal index depends on three superconformal fugacities $(p, q, t)$
and on any number of fugacities $\{ a_i\}$ associated to
flavor symmetries (which, by definition, commute with the superconformal algebra),\footnote{In this review we follow the conventions of~\cite{Beem:2013sza}.  In comparing with \cite{Gadde:2011uv, Gaiotto:2012xa}, 
the only significant change is $j_1 \to -j_1$ in the definitions of $j_{12}$ and $j_{34}$. The conventions for labeling supercharges 
 are also slightly different in these two sets of references,
 but  notations aside  all of them choose  ``same'' supercharge  to define the general index ({\it i.e.} the supercharge with quantum numbers $E=R=-r = \half$, 
 $(j_1, j_2) = (0, -\half)$. 
}
\begin{equation}\label{inddef}
{\cal I} (p, q, t ; a_i) \colonequals \mathrm{Tr} (-1)^F\,\left(\frac{t}{pq}\right)^r\,
 p^{j_{12}}\,
 q^{j_{34}}\,
 t^{R}\,
 \prod_i a_i^{f_i} \,  e^{-\beta \delta_{2 \dot -}} \, , 
\end{equation}
where
\be\label{BPS}
2 \delta_{2 \dot -} \colonequals \{\widetilde {\cal Q}_{2\dot -},\, \widetilde {\cal Q}_{2\dot -} ^\dagger \}=
E-2j_2-2R+r \,. 
\ee 
We will always assume that
\bea
|p|<1\,,\qquad
|q|<1\,,\qquad
|t|<1\,,\qquad 
|a_i|= 1\,,\qquad
\left|\frac{p\,q}{t}\right|<1\,.
\eea
Our notations are as follows. We denote by $E$ the conformal hamiltonian (dilatation generator), by
$j_1$ and $j_2$ the Cartan generators of 
of the $SU(2)_1\times SU(2)_2$  isometry group of $\S^3$, by $R$ and $r$ the Cartan generators of the
  $SU(2)_R \times U(1)_r$ the superconformal R-symmetry.
We have also defined
 $j_{12} \colonequals j_2- j_1$ and $j_{34} \colonequals j_2+j_1$,
 which generate rotations in two orthogonal planes (thinking of ${\mathbb S}^3$ as embedded in ${\mathbb R}^4$).
   Finally  $\{ f_i \}$ are the flavor symmetry generators. In our conventions, we
  label the supercharges as 
  \be
  {\cal Q}^{\cal I}_\alpha \, \qquad \tilde {\cal Q}_{ {\cal I} \dot \alpha}\, \qquad {\cal S}_{\cal I}^\alpha \,  \qquad \tilde {\cal S}^{ {\cal I} \dot \alpha}  \, ,
  \ee
where $\alpha = \pm$ is an $SU(2)_1$ index, $\dot \alpha = \dot \pm$ an $SU(2)_2$ index and
 ${\cal I}=1, 2$ an $SU(2)_R$ index. We have 
 ${\cal S} = {{\cal Q}}^\dagger$ and $\tilde {\cal S} = { \tilde {\cal Q}}^\dagger$.
Writing an explicit trace formula  for the index involves a choice of supercharge.
With no loss of generality, we chose in ~\eqref{inddef}  to count cohomology classes
of $\tilde {\cal Q}_{2\dot -}$, which has quantum numbers
$E= R= -r = \half$,  $(j_1, j_2) = (0, -\half)$. 
The states  that to this index are the ``harmonic representatives'' satisfy $\delta\equiv \delta_{2 \dot -} = 0$.
All other choices of a Poincar\'e supercharge
would  give an equivalent index~\cite{Kinney:2005ej}.

In appendix \ref{shorteningappendix} we review the shortening conditions of the ${\cal N}=2$ superconformal algebra
and the recombination rules of short multiplets into long ones. Explicit formulae for the index
of individual short multiplets are given in appendix B of \cite{Gadde:2011uv} and will not be repeated here.
It is important to keep in mind that knowledge of the index alone is in general not sufficient to completely reconstruct
the spectrum of short representations of a given theory. Schematically, the issue is the following \cite{Gadde:2009dj}.
Suppose that two short multiplets, $S_1$ and $S_2$,
can recombine to form a long multiplet $L_1$, 
\be
S_1 \oplus S_2 = L_1 \, ,
\ee
and similarly that $S_2$ can recombine with a third short multiplet $S_3$ to give another long multiplet $L_2$,
\be
S_2 \oplus S_3 = L_2\,.
\ee
By construction, the index evaluates to zero on long multiplets, so
\be
\II(S_1) = - \II(S_2) = \II(S_3) \,.
\ee
The index cannot distinguish between the two multiplets $S_1$ and $S_3$.
(Note that  $S_2$ {\it is} distinguished from $S_1 \sim S_3$ by  the overall sign.) 
A detailed discussion of equivalence classes of  multiplets 
that have the same ${\cal N}=2$ superconformal index  can be found in 
section 5.2 of \cite{Gadde:2009dj}.

\

\subsection{Free field combinatorics}

The simplest examples of conformal quantum field theories are  free theories.
In a free theory, the general local operator is obtained from normal ordering of the elementary fields,
and its quantum numbers including the conformal dimension take their classical ``engineering'' values.
By the state/operator map, local operators inserted at the origin are in one-to-one correspondence with states.
Enumerating states reduces then to the simple combinatorial problem of enumerating
all possible composite ``words'' (or ``multi-particles'') built out of  the elementary ``letters'' (or ``single-particles''),
which are the elementary fields 
and their space-time derivatives.

For our purposes, we are interested in enumerating states with $\delta=0$, and since in a free theory the value
of $\delta$ of a composite operator is simply the sum of the values of $\delta$ of its elementary letters, we may from the start
restrict to the letters with $\delta =0$. The letters contributing the index of the free ${\cal N}=2$  hypermultiplet and of the free 
 vector multiplet 
 ${\cal N}=2$ are shown in Table~\ref{letters}. One immediately finds 
the following single-particle indices ({\it i.e.}, the indices computed over the set of single-particle states):
\bea \label{spindices}
&&{\cal I}^{s.p.}_H=t^{\frac{1}{2}}\,\frac{1-\frac{p\,q}{t}}{(1-p)(1-q)}(a+a^{-1})\,\chi_\Lambda(\bf x)\,,\\
&&{\cal I}^{s.p.}_V=-\frac{q}{1-q}-\frac{p}{1-p}+\frac{\frac{p\,q}{t}-t}{(1-p)(1-q)}\,.
\eea  
Here $a$ is a $U(1)$ fugacity under which the two half-hypers have opposite charges and $\chi_\Lambda(\bf x)$ 
is the character of the representation of some global symmetry.
The multi particle-indices are given by the plethystic exponentials of the single-particle ones. In 
particular the index of a free hypermultiplet in a bi-fundamental representation of $SU(n)\times SU(n)$, which will play an important role in our discussion, 
is given by
\bea \label{hyperindex}
{\cal I}_H(a, {\bf x}, {\bf y}; p, q, t) &=&{\rm PE}\left[t^{\frac{1}{2}}\,\frac{1-\frac{p\,q}{t}}{(1-p)(1-q)}(a+a^{-1})
\left(\sum_{i=1}^n x_i\right)\left(\sum_{j=1}^n y_j\right)\right]\nonumber\\
&=&\prod_{i,j=1}^n \Gamma(t^{\frac12}\,a\,x_i\,y_j\,;\,p,\,q)\,\Gamma(t^{\frac12}\,(a\,x_i\,y_j)^{-1};\,p,\,q)\,.
\eea We collect the definitions of the plethystic exponenential, elliptic Gamma function, and related  objects 
 in appendix~\ref{notationssec}.
\begin{table}
\begin{centering}
\begin{tabular}{|c|r|r|r|r|r|c|c|}
\hline
Letters & $  E$ & $j_1$ & $  j_2$ & $R$ & $r$    & $\mathcal{I}(p, q, t)$ \tabularnewline
  \hline
   \hline
$  \phi$ & $1$ & $0$ & $0$ & $0$ & $-1$   & $pq/t$   \tabularnewline
  \hline
$  \lambda^{1} _{\pm}$ & $  \frac{3}{2}$ & $  \pm  \frac{1}{2}$ & $0$ & $  \frac{1}{2}$ & $-  \frac{1}{2}$   &  $-p$, $-q$ \tabularnewline
  \hline
$  \bar{\lambda}_{2 \dot{+}}$  & $  \frac{3}{2}$ & $0$ & $  \frac{1}{2}$ & $  \frac{1}{2}$ & $  \frac{1}{2}$  &  $-t$ \tabularnewline
  \hline
$  \bar{F}_{\dot{+}\dot{+}}$ & $2$ & $0$ & $1$ & $0$ & $0$   &  $pq$ \tabularnewline
  \hline
  $  \partial_{-\dot{+}}  \lambda^1_{+}+  \partial_{+\dot{+}}  \lambda^1_{-}=0$ & $  \frac{5}{2}$ & $0$ & $  \frac{1}{2}$ & $  \frac{1}{2}$ &
 $  -\frac{1}{2}$ & $pq$  \tabularnewline
  \hline
\hline
$Q$ & $1$ & $0$ & $0$ & $  \frac{1}{2}$ & $0$  &  $\sqrt{t}$ \tabularnewline
  \hline
$  \bar{\psi}_{\dot{+}}$ & $  \frac{3}{2}$ & $0$ & $  \frac{1}{2}$ & $0$ & $-  \frac{1}{2}$   & $-pq/\sqrt{t}$ \tabularnewline
  \hline
    \hline
$  \partial_{\pm\dot{+}}$ & $1$ & $  \pm  \frac{1}{2}$ & $  \frac{1}{2}$ & $0$ & $0$    & $p$, $q$ \tabularnewline
\hline
\end{tabular}
\par  \end{centering}
  \caption{Contributions to the index from  single-particle (letter) operators of the two basic ${\cal N}=2$ multiplets: the vector multiplet and the hypermultiplet.
 We denote by $(\phi, \bar \phi,  \lambda^{\cal I}_{\alpha}, \bar\lambda_{{\cal I}\,\dot \alpha},  F_{\alpha \beta}, \bar F_{\dot \alpha \dot \beta})$
 the components of the adjoint ${\cal N} = 2$ vector multiplet,  by $(Q, \bar Q, \psi_\alpha, \bar \psi_{\dot \alpha})$ the
 components  of the  ${\cal N} = 1$
chiral multiplet,   and by $\partial_{\alpha \dot \alpha}$ the spacetime derivatives.
}
\label{letters}
\end{table}

\

Conversely, one of  the hallmarks of a free theory is the  fact
that the plethystic log of the index is simple.
For example, formally analogous to the counting problem in free field theory is the counting problem
for large $N$ theories. It often happens that the conformal gauge theories come in families labeled by the rank
of the gauge group and in the limit of large rank they have a dual description in terms
of supergravity in $AdS$ backgrounds~\cite{Kinney:2005ej}. In such cases the operators counted by the index are dual to free supergravity modes. Thus, taking the limit of large $N$ the index reduces again to a simple plethystic exponential
of the towers of single trace operators dual to the finite number of free supergravity fields.

\

\subsection{Gauging}

After we dealt with free  theories, let us turn to interacting models.  In general, we should not expect any simple
combinatorial description of the set of local operators in an interacting theory.
An important exception are the 
superconformal field theories that admit a Lagrangian description,
which by definition are continuously connected to free field theories by turning off the gauge
couplings. 
Since the index is independent of exactly marginal deformations, we may as well compute
it in the free limit (setting to zero all gauge couplings). The only effect of the gauging is the 
Gauss law constraint, {\it i.e.} the  projection
onto gauge invariant states. 

More generally, starting from a SCFT ${\cal T}$,  we can obtain a new superconformal field theory ${\cal T}_G$
by gauging a subgroup $G$ of the flavor symmetry of ${\cal T}$, provided of course that the gauge coupling beta functions
vanish. 
If the index of ${\cal T}$ is known, we find the index of ${\cal T}_G$ by multiplying by the index
of a vector multiplet in the adjoint representation of $G$, and then integrating over $G$  
 with the invariant Haar measure to enforce the projecting over gauge singlets,
\be \label{generalgauging}
{\cal I} [ {\cal T}_G] =\int [d{\bf z}]_G\; {\cal I}_V({\bf z})\;{\cal I}[ {\cal T} ]({\bf z})\,.
\ee 
 In fact we can treat the index ${\cal I}_{ {\cal T} }({\bf z})$ as a ``black-box'': it might be the
index of a collection of free hypermultiplets, the index of a gauge theory, 
or the index of an interacting theory for which we do not know a useful description in terms of a Lagrangian. 
Whenever a flavor symmetry is gauged in four dimensions, the effect on the index is 
simply to introduce the vector multiplet and project onto gauge-invariant states.\footnote{
In other dimensions the situation can be slightly more involved. For example, in three dimensions
a gauge theory contains local monopole operators which have to be introduced into the index computations along with the vector multiplets.
}

In all known examples,  conformal manifolds of ${\cal N}=2$ SCFTs are parametrized by gauge couplings.
It is tempting to speculate that the most general ${\cal N}=2$ SCFT is obtained by gauging a set of elementary
building blocks, each of which is an isolated theory with no exactly marginal couplings.  The simplest
 of such an elementary building block is the free hypermultplet theory. We will encounter below
several other examples of building blocks with no known Lagrangian description. Determining the index of such
isolated theories would appear to be very challenging. Fortunately, for theories of class ${\cal S}$ we can leverage
the additional structure of generalized S-duality.
Let us turn to a concrete illustration.

\

\section{Interlude: duality and the index of $E_6$ SCFT}\label{E6}

In this section, we will sketch how to determine the index of a canonical example of isolated non-Lagrangian theory,
the  SCFT with $E_6$ flavor symmetry of Minahan and Nemeschansky~\cite{Minahan:1996fg}. The general idea is to couple
the isolated theory to some extra stuff, and use dualities to relate the larger theory to a more tractable model.

For the case at hand, we exploit Argyres-Seiberg duality~\cite{argyres-2007-0712}.
On one side of the duality we have an $SU(3)$ SYM with $N_f=6$ flavors.
On the other side of the duality we have a hypermultiplet in the fundamental representation
of gauged $SU(2)$ under which also a strongly-coupled theory with $E_6$ flavor symmetry~\cite{Minahan:1996fg}
 is charged. The $SU(2)$ gauged group is a sub-group of the $E_6$ flavor symmetry.
By the  rules of computing the index reviewed in the previous section, this duality can be written as equality of two integrals~\cite{Gadde:2010te},
\bea\label{E6int}
&&\int [d{\bf z}]_{G=SU(3)}\, {\cal I}^{SU(3)}_V({\bf z})\,
{\cal I}^{(1)}_H({\bf z},{\bf y},{\bf x},a,b)=\\
&&\qquad\qquad\qquad\int [d{z}]_{G=SU(2)}\, {\cal I}^{SU(2)}_V(z)\,{\cal I}^{(2)}_H(z,(a/b)^{3/2})\,
{\cal I}_{E_6}({\bf x},{\bf y},\{z (ab)^{-\half},z^{-1}(ab)^{-\half}\})\,.\nonumber
\eea Here ${\cal I}_{E_6}({\bf x}_1,{\bf x}_2,{\bf x}_3)$ is the unkown index of the theory with $E_6$ flavor symmetry with $\{{\bf x}_i\}$ being the fugacities for $SU(3)^3$ maximal subgroup of $E_6$;
${\bf y}$ and ${\bf x}$ are $SU(3)$ fugacities and $a,b$ are two $U(1)$ fugacities. The quantity
${\cal I}^{(1)}_H({\bf z},{\bf y},{\bf x},a,b)$ represents the index of a collection of hypermultiplets in the bi-fundamental representation
of flavor of $SU(3)^2$ and the gauged $SU(3)$, whereas 
${\cal I}^{(2)}_H(z,(a/b)^{3/2})$ is a fundamental hypermultiplet of $SU(2)$.
The powers of $U(1)$ fugacities $a$ and $b$ on the right-hand side  of the equality are a consequence
of the details of the map of global symmetries between the two duality frames.
 In general from equalities
of integrals of this sort one cannot extract the precise values of the integrands. However, in this particular
case the integral on the right-hand side is invertible and just by assuming the Argyres-Seiberg 
duality as manifested for the index in~\eqref{E6int} one can explicitly deduce the index ${\cal I}_{E_6}$.
Schematically, this inversion procedure takes the following form
\be\label{E6sol}
{\cal I}_{E_6}({\bf x},{\bf y},\{c (ab)^{-\half},c^{-1}(ab)^{-\half}\})=
\oint_{\cal C}\frac{d\,h}{2\pi i h} \Delta(h, c)\,
\int [d{\bf z}]_{G=SU(3)}\, {\cal I}^{SU(3)}_V({\bf z})\,
{\cal I}^{(1)}_H({\bf z},{\bf y},{\bf x},a,b)\,.\nonumber\\
\ee
Here ${\cal C}$ is a well-defined integration contour and $\Delta$ is a specific inversion kernel~\cite{spirinv}. 
Physically, the fact that the integral is invertible means that the extra hyper-multiplet introduced
while gauging a sub-group of the $E_6$ symmetry adds
enough structure so that the information about
 the protected spectrum of the $E_6$ theory itself, a-priori lost after gauging,
can be still recovered.

We thus are able to completely fix the superconformal index of a theory not connected to a free theory
by a continuous parameter. The trick is to {\it enlarge} the theory with the bigger theory admitting 
an alternative description which {\it can} be connected to a free theory by continuous deformation.
This basic idea will be behind the general procedure we will outline in the next sections.

\

Before turning to the general discussion of class ${\cal S}$ theories, let us 
illustrate in this concrete example what kind of physical information can be extracted from the index.
 Explicitly computing~\eqref{E6sol} one obtains to the lowest orders in the series expansion in fugacities
\bea
{\cal I}_{E_6}&=&PE[{\cal I}_u]\,PE[{\cal I}_H(\chi_{78})]\,PE[{\cal I}_T]\;\times\\
&&\qquad \left(1-(t^2-p\,q\,t+t^2(p+q))\,(\chi_{650}+1)-\frac{p^3q^3}{t^2}+p\,q\,t\,\chi_{78}+\cdots\right)\,.\nonumber
\eea In the first line we have the protected multiplets appearing 
in this theory: ${\cal I}_u$ is the Coulomb branch multiplet (the dual of $u=Tr \phi^3$
of the $SU(3)$ gauge theory), ${\cal I}_H$ is the Higgs branch generator, $X$,  in ${\bf 78}$
of the $E_6$ global symmetry, and ${\cal I}_T$ is the stress-energy multiplet. The quantum numbers 
of these multiplets are different from free fields. Moreover, on the second line we have {\it constraints}
appearing removing some of the contributions generated on the first line: these constraints are the footprint of the non-trivial dynamics of the theory.
 For example one constraint encoded here is
\be\label{Jos}
\left[X\otimes X\right]_{650\oplus 1}=0\,,
\ee which is the Joseph's relation discussed in~\cite{Gaiotto:2008nz}.

\

\section{Derivation of the index for theories of class ${\cal S}$}\label{secDiff}

In this section we will determine the index for all theories of class ${\cal S}$. Broadly speaking, we will be
 using the same kind of physical input as in the previous section, namely   knowledge of the index for Lagrangian theories and the assumption of generalized S-duality.
 We will  however exploit these ingredients in a different way, arriving at a particularly elegant and uniform description
of the general index. 
For simplicity we focus on the basic index (the $\S^3 \times \S^1$ partition function),
and to the simplest class ${\cal S}$ theories of type $A$. Several generalizations will be mentioned in section \ref{S:generalizations}.

\subsection{Class ${\cal S}$}

A lightening review of class ${\cal S}$ is in order. A  $4d$ superconformal field theory of  class ${\cal S}$ is specified
 the following data:\footnote{These are the ``basic'' theories. A larger list is obtained by allowing
 for ``irregular'' punctures. Further possibilities arise by decorating the UV curve with outer automorphisms twist lines , see \cite{Tachikawa:2010vg}.
} 
\begin{itemize}
\item
A choice of  the type  $\gf$ of the $(2, 0)$ theory, where $\gf = \{ A_n, D_n, E_6, E_7, E_8\}$
 is a simply-laced Lie algebra.
\item A choice of  UV curve ${\cal C}_{g, s}$, where $g$ indicates the genus and $s$ the number of punctures of the curve. 
Only the complex structure moduli of ${\cal C}_{g, s}$ matter. They are interpreted as the exactly marginal gauge
couplings of the $4d$ SCFT.
\item Each puncture corresponds to a codimension two defect of the $(2, 0)$ theory. We restrict to the so-called  regular
defects, which are 
labelled by a choice of embedding $\Lambda: \suf (2) \to \gf$.  
The centralizer
$\hhf \subset \gf$ of the image of $\Lambda$ in $\gf$ is the flavor symmetry associated to the defect.
All in all,
the theory enjoys at least\footnote{In some special cases, the symmetry is enhanced by 
 additional generators which are not naturally assigned to any puncture. 
}
 the flavor symmetry algebra $\oplus_{i=1}^s\hhf_i$. \end{itemize}
We will label the corresponding $4d$ SCFT as ${T}[\gf; {\cal C}_{g, s}; \{ \Lambda_i \} ]$. 

From now on we will restrict our discussion to class  ${\cal S}$ theories of type  $A$,  $\gf = A_{n-1}$.  
The embeddings $\Lambda: {\suf (2)} \to \suf(n)$
are in one-to-one correspondence with partitions of $n$, $[n_1^{\ell_1}, n_2^{\ell_2}, \dots n_k^{\ell_k}]$ with $\sum_i \ell_i n_i = n$ and $n_i  > n_{i+1}$,
which indicate how the fundamental representation
of $\suf(n)$ decomposes under representations of $\Lambda (\suf (2))$.
For the trivial embedding $\Lambda = 0$,  associated to the partition $[1^n]$,
we have maximal flavor symmetry $\hhf = \suf(n)$ and the corresponding
puncture is called {\it maximal}. The other extreme case case is the principal embedding, associated to the partition $[n]$,
leading to $\hhf = 0$ (no flavor symmetry),
so the puncture is effectively deleted. 
Another important case is the subregular embedding, associated to the partition $[n-1, 1]$,
which leads to $\hhf = \uf (1)$, the smallest non-trivial flavor symmetry, so the
 corresponding puncture is called a {\it minimal} puncture.\footnote{
Throughout this review we will often associate punctures with flavor symmetry factors.
For theories of type $A$ this association is well motivated (although there can be two different punctures with same flavor symmetry), but one has to remember that for type $D$ and $E$ theories one can have non-trivial 
punctures with no flavor symmetry associated with them.
}

The surface ${\cal C}_{g, s}$ can be assembled
by gluing together three-punctured spheres, or ``pairs of pants'' (viewed as three-vertices) and cylinders (viewed as propagators).
Each cylinder is associated to a simple gauge group factor of the $4d$ SCFT, with 
the plumbing parameter interpreted as the corresponding marginal gauge coupling. The degeneration limit
of the surface where one cylinder becomes very long corresponds to the weak coupling limit of that gauge group.
Cutting a cylinder is interpreted as
``ungauging'' an $SU(n)$ gauge group, leaving  behind two maximal punctures, each carrying
$SU(n)$ flavor symmetry. Conversely, gluing  two maximal punctures corresponds to gauging the diagonal subgroup
of their $SU(n) \times SU(n)$ flavor symmetry. 
The basic building blocks of class ${\cal S}$ are thus the theories associated to three-punctured spheres,  $T_n^{\Lambda_1, \Lambda_2, \Lambda_3} \colonequals
{T} [ \suf(n); {\cal C}_{0, 3} ; \Lambda_1\, \Lambda_2\, \Lambda_3 ]$. These are isolated SCFTs with no tunable couplings, in harmony with the fact that three-punctured
spheres carry no complex structure moduli. Most of them have no known Lagrangian description. An important exception
is the theory associated to two maximal and one minimal puncture, $T_n^{[1^n]\, [1^n]\, [n-1, 1]}$,
which is identified with the free hypermultiplet
in the bifundamental representation of $SU(n) \times SU(n)$.

\begin{table}[h]
\centering
\begin{tabular}{|@{$\Bigm|$}c|c@{$\Bigm|$}|}
\hline
\textbf{ $4d$ theory   $T[{\cal C}]$  }  & \textbf{ Riemann surface  ${\cal C}$ }\\
\hline
\hline
Conformal manifolds & Complex structure moduli of ${\cal C}$\\
\hline
$SU(n)$ gauge group &  cylinder \\
with coupling $\tau$ & with sewing parameter $q=\exp (2\pi i \tau)$ \\
\hline
Flavor-symmetry factor   & Puncture labelled by $SU(2) \to SU(n)$ \\
$H \subset SU(n)$ & with commutant $H$ \\
\hline
Weakly-coupled frame & Pair-of-pant decomposition of ${\cal C}$ \\
\hline
Generalized $S$-duality & Moore-Seiberg groupoid of ${\cal C}$ \\
\hline
\hline
Partition function on $S^4$ & Correlator in Liouville/Toda on ${\cal C}$\\
\hline
\hline
Superconformal index & Correlator in a TQFT on ${\cal C}$\\
\hline
\end{tabular}
\caption{ The basic class ${\cal S}$ dictionary.
}
\end{table}

Different  pairs-of-pants decompositions of the UV curve correspond  to different weakly coupled descriptions of the same SCFT,
related by generalized S-dualities. The Moore-Seiberg groupoid of the UV curve is thus identified with the S-duality groupoid
of the SCFT.

\subsection{TQFT interpretation of the index}

The index of  ${T}[\gf; {\cal C}_{g, s}; \{ \Lambda_i \} ]$ is a function of the superconformal fugacities $(p, q, t)$ and of the
flavor fugacities ${\bf a}_i$, $i=1, \dots s$, associated to the Cartan generators of the global symmetry group $H_1 \otimes \dots \otimes H_s$,
but it is independent of the complex structure moduli of the UV curve ${\cal C}_{g, s}$.
We can thus  regard the index as a correlator of a TQFT defined on the UV curve~\cite{Gadde:2009kb},
\be
{\cal I}^g [ p, q, t; {\bf a}_i] = \langle  {\cal O} ({\bf a}_1)  \dots {\cal O} ({\bf a}_s)  \rangle_{ {\cal C}_{g, s} } \, ,
\ee
where we have formally introduced ``local operators'' ${\cal O}({\bf a}_i)$ associated to the punctures. This is natural, because
the index enjoys the kind of factorization  property expected for a TQFT correlator. Given  a pair-of-paints decomposition
of ${\cal C}_{g, s}$ we may cut an internal cylinder and disconnect the surface into the two surfaces\footnote{This is the generic situation.
The remaining possibility is that cutting the cylinder yields the connected surface ${\cal C}_{g-1, s+2}$.
This case can be treated analogously.}
 ${\cal C}_{g_1, s_1 +1}$
and ${\cal C}_{g_2, s_2 +1}$, with $g_1 + g_2 = g$ and $s_1 + s_2 = s$. By applying the general gauging prescription (\ref{generalgauging}),
we have the ``factorization'' formula\footnote{We'll often omit the dependence on the superconformal fugacities to avoid cluttering.}
\be
{\cal I}^g [ {\bf a}_1, \dots  {\bf a}_s ]  = \int [d{\bf b}]_G \;   \; {\cal I}^{g_1} [ {\bf a}_j, {\bf b} ]   \; {\cal I}_V({\bf b}) \; {\cal I}^{g_2} [{\bf b}, {\bf a}_k]  \, , \quad   j \in S_1, \, k \in S_2\, ,
\ee
where $S_1$ and $S_2$ are the set of indices labeling the punctures on the two components, with $S_1 \cup S_2 = \{ 1, \dots s\}$.
As the index is 
invariant under generalized S-dualities, one must obtain the same answer by applying the factorization formula in different channels.
This is the essential property that must be satisfied by a $2d$ TQFT correlator.

To make the connection with the standard treatment of $2d$ TQFT more explicit, let us make a change
of basis, from a continuous to a discrete set of operators. For simplicity we restrict to the case
where all punctures are maximal, carrying the full flavor symmetry $\gf$. The operator
${\cal O}( \bf a)$ is labelled by the flavor  fugacity $\bf a$ dual to the Cartan subalgebra of $\gf$. Consider
now a complete set of Weyl invariant functions $\{ \psi_\alpha (\bf a) \}$, where the label $\alpha$ runs over the finite-dimensional
irreps of $G$, and define the discrete set of operators ${\cal O}_\alpha$
by the integral transform
\be
{\cal O}_\alpha \colonequals  \int  [d{\bf a}]_G \, {\cal I}_V ({\bf a})\, \psi_\alpha ( \bf a) {\cal O}( \bf a)\,.
\ee
It is convenient to choose the $\{ \psi_\alpha (\bf a) \}$ to be orthonormal under the propagator measure,
\be \label{propmeasure}
\int  [d{\bf a}]_G  \, {\cal I}_V({\bf a})  \psi_\alpha (\bf a)  \psi_\beta (\bf a) = \delta_{\alpha \beta} \,.
\ee
In this discrete basis, the factorization property reads simply
\be
{\cal I}^g_{ \alpha_1, \dots  \alpha_s }  =  {\cal I}^{g_1}_{ \{\alpha_j\} \, \beta}  \;  {\cal I}^{g_2}_{\beta \,\{ \alpha_k \}} \, , 
\ee
where the repeated index $\beta$ is summed over. It is then clear that the general correlator on an surface of arbitrary topology
can be obtained by successive contractions of the  three-point correlator, {\it i.e.} the index of the three-punctured sphere, 
 ${\cal I}^{g=0}_{\alpha_1 \alpha_2 \alpha_3} \equalscolon C_{\alpha_1 \alpha_2 \alpha_3}$.
 These ``TQFT structure constants''  $C_{\alpha_1 \alpha_2 \alpha_3}$ are symmetric functions  of the three labels $\alpha_i$
 and must satisfy the associativity constraint 
 that follows from demanding that factorization of ${\cal I}^{g=0}_{\alpha_1 \alpha_2 \alpha_3 \alpha_4}$
 in two different ways must yield the same result,
\be \label{ass}
C_{\alpha_1\alpha_2\beta } \; C_{\beta \alpha_3\alpha_4}=
C_{\alpha_1\alpha_3\gamma}\; C_{\gamma \alpha_2\alpha_4}\,.
\ee 
This condition is in fact sufficient to ensure  independence of the general correlator on any specific choice of pair-of-pants decomposition.
The structure that we have just described is very close to the standard axiomatic description of $2d$ TQFTs, but with the caveat
that in the mathematical literature the state-space of the TQFT is usually taken to be finite-dimensional,
whereas we have the infinite-dimensional space of finite-dimensional irreps of~$\gf$.

It is a simple linear algebra fact that one may always\footnote{Here we should mention  that since the state-space of the QFT obtained from the index is infinite dimensional there might be in principle issues of converges when changing basis. Such complication though do not actually arise in practice in the index computations.} perform a further change of basis
to a preferred discrete basis, in which
associativity
relations~\eqref{ass}  become trivial (see appendix~A of~\cite{Gadde:2011uv} for an explicit example). This is the so-called Frobenius basis,
which is still orthonormal under the propagator measure  and is such
that the structure constants have the  diagonal  structure
\be
C_{\lambda_1\lambda_2\lambda_3}=C_\lambda \,
\delta_{\lambda\lambda_1}\,
\delta_{\lambda\lambda_2}\,
\delta_{\lambda\lambda_3}\,.
\ee 
In the Frobenius basis the non-vanishing components of the index associated to ${\cal C}_{g, s}$ take the
very simple form
\be
{\cal I}^g_{ \lambda \dots  \lambda } =   C_\lambda^{2g -2 + s} \, ,
\ee
which just follows from the observation that  ${\cal C}_{g, s}$ can be built by gluing $(2g-2+s)$ three-punctured sphere,
and that the contractions of indices implementing the gluings are all trivial in this basis.
Going back to the continuous fugacity basis,
\be \label{Ipsi}
{\cal I}^g [ {\bf a}_1, \dots  {\bf a}_s ] =  \sum_\lambda   C_\lambda^{2g -2+ s} \,  \psi_\lambda ( {\bf a}_1) \dots\psi_\lambda ( {\bf a}_s)\,.
\ee
In summary, the task of evaluating the general index is reduced to the task of finding the Frobenius basis $\{ \psi_\lambda (\bf a)\}$
and the structure constants $C_\lambda$.


\subsection{Bootstrapping the index}

The  structure just outlined is so constraining
that it  essentially fixes the index of class ${\cal S}$ theories,
when supplemented with the extra physical input about the special cases that have a 
Lagrangian description~\cite{Gaiotto:2012xa}.

We focus on $A_{n-1}$ theories. Let us first aim to find the index for theories containing only maximal
punctures.  For $n >2$, none of these theories have a Lagrangian description.
Nevertheless, their index must obey compatibility conditions that follow
by gluing in an extra three-punctured sphere of type
$T_n^{[1^n]\, [1^n]\, [n-1, 1]}$, which is 
 identified with the free hypermultiplet theory in the bifundamental of $SU(n) \times SU(n)$.
The  physical input  mentioned above  is then
\be
{\cal I} [ T_n^{[1^n] \,[1^n]\, [n-1, 1]} ]  = {\cal I}_H (a, {\bf x}, {\bf y} ) \, ,
\ee
where the explicit expression of  ${\cal I}_H$ is given in (\ref{hyperindex}). Recall that
 $a$ is the $U(1)$ fugacity associated with minimal puncture while ${\bf x}$, ${\bf y}$ the $SU(n)$ fugacities associated with the two maximal punctures.

Let the index of  $T[ {\cal C}]$ with all maximal punctures be some unknown function\footnote{The dependence on the superconformal fugacities
$(p, q, t)$ is again left implicit.} 
${\cal I}_{\cal C}( {\bf x}_i)$,
symmetric under permutations of the arguments ${\bf x}_i$, $i = 1, \dots s$.
We construct a larger theory  with $s$ maximal and one minimal
puncture by gluing in a free hypermultiplet. The resulting index is given by
\be\label{frameA}
{{\cal I}}(a,{\bf x}_1, {\bf x}_2 \cdots {\bf x}_s)=\int [d{\bf z}] \;{\cal I}_V({\bf z})\,{\cal I}_H(a, {\bf x}_1, {\bf z})\, {\cal I}_{\cal C}({\bf z}^{-1},{\bf x}_2 ,\cdots, {\bf x}_s)\,.
\ee  
While in the above expression ${\bf x}_1$ appears to be treated asymmetrically from ${\bf x}_2, \dots  {\bf x}_s$, 
generalized S-duality (the TQFT structure of the index) demands that the integral be invariant under permutations
of all the ${\bf x}_i$.  Remarkably, this will be sufficient
to determine the function ${\cal I}_{\cal C}$. To reach this conclusion,
we take an apparent detour and study the analytical properties of the integral as a function of the $U(1)$ fugacity $a$.

One can show that the
integral has simple poles for
\be\label{apoles}
p^{r} q^{s}   t^{\frac{n}{2}} a^{- n}  = 1\, ,
\ee
where $r$ and $s$  non-negative integers.
To see this one notices that the poles in ${\bf z}$ in the integrand move around when one varies 
$a$. At the special values~\eqref{apoles} pairs of poles pinch the integration contours and cause the whole integral to diverge.
  A toy example of this mathematical phenomenon is as follows. Consider ($|t\,a|,\,|t\,b|,\,|t\,c|<1$)
$$\oint\frac{dz}{2\pi i z}\oint \frac{dy}{2\pi i y}\frac{1}{t\, a- y}\frac{1}{t\, b- z} \frac{1}{t^{-1} c^{-1}- z\,y}=
\oint\frac{dz}{2\pi i z}\frac{1}{t\, b- z} \frac{1}{t^{-1} c^{-1}- z\,t\,a}=\frac{t\,c}{1- t^{3}\,a\, b\,c}\,.
$$ We have a pole at $t^3a\,b\,c=1$. 
This can be viewed as the pole in $y$ at $t\,a$ colliding with pole in $y$
at $t^{-1}c^{-1}z^{-1}$ simultaneously with pole in $z$ at  $t\,b$ colliding with the pole at 
$t^{-1}c^{-1}y^{-1}$.

The residues of the poles~\eqref{apoles} are easy to compute. This residue gets contributions
in the ${\bf z}$ contour integrals only from the finite number of poles that pinch the integration
 contours.  
The simplest case is the residue at $ t^{\frac{n}{2}} a^{-n} =1$,
\be\label{basres}
{\cal I}_V\, {\rm Res}_{t^{\frac{n}{2}} a^{-n}   \to 1} {\cal I}(a,{\bf x}_1,{\bf x}_2,\cdots)= {\cal I}_{\cal C}({\bf x}_1,{\bf x}_2,\cdots)\, ,
\ee
where ${\cal I}_V$ is the index of $U(1)$ ${\cal N}=2$ vector multiplet. So picking
up the residue at $a^2 t^{-1} =1$ has the effect of ``deleting'' the extra $U(1)$ puncture.
A slightly more involved calculation gives the residue  at $ q^{} t^{\frac{n}{2}} a^{-n} = 1$,
\bea\label{basicOp}
&&{\cal I}_V\, {\rm Res}_{ q t^{\frac{n}{2}} a^{-n} \to 1 } {\cal I}(a,{\bf x},{\bf y},\cdots)=\\
&&\qquad\qquad= \frac{\theta(t;p)}{\theta(q^{-1};p)}\,\sum_{i=1}^n
\prod_{j\neq i}\frac{\theta(\frac{t}{q}x_i/x_j;p)}{\theta(x_j/x_i;p)}\,
{\cal I}_{\cal C}(\{x_i\to q^{-\frac{1}{2}}x^{}_i,\, x_{j\neq i}\to q^{\frac{1}{2}}x_j^{}\},{\bf y},\cdots)\, \nonumber\\
&&\qquad\qquad \equalscolon \frac{\theta(t;p)}{\theta(q^{-1};p)}\,{\frak S}_{(r=0,s=1)}({\bf x})\,{\cal I}_C({\bf x},{\bf y},\cdots)\, .\nonumber
\eea 
We see that the residue is computed by the action on ${\cal I}_C$ of an interesting
  difference operator, which we have named   ${\frak S}_{(r=0,s=1)}({\bf x})$, shifting the values of the fugacity ${\bf x}$.   
  The residues can be easily computed for general values of $r$ and $s$ in~\eqref{apoles},
and are again given by acting  on ${\cal I}_C$ with certain difference operators  ${\frak S}_{(r,s)}({\bf x})$ which we will not write explicitly.
The operators ${\frak S}_{(r,s)}({\bf x})$ 
all commute with each other
 and are self-adjoint under the propagator measure.

\begin{figure}[ht]
\begin{center}
\includegraphics[scale=.6]{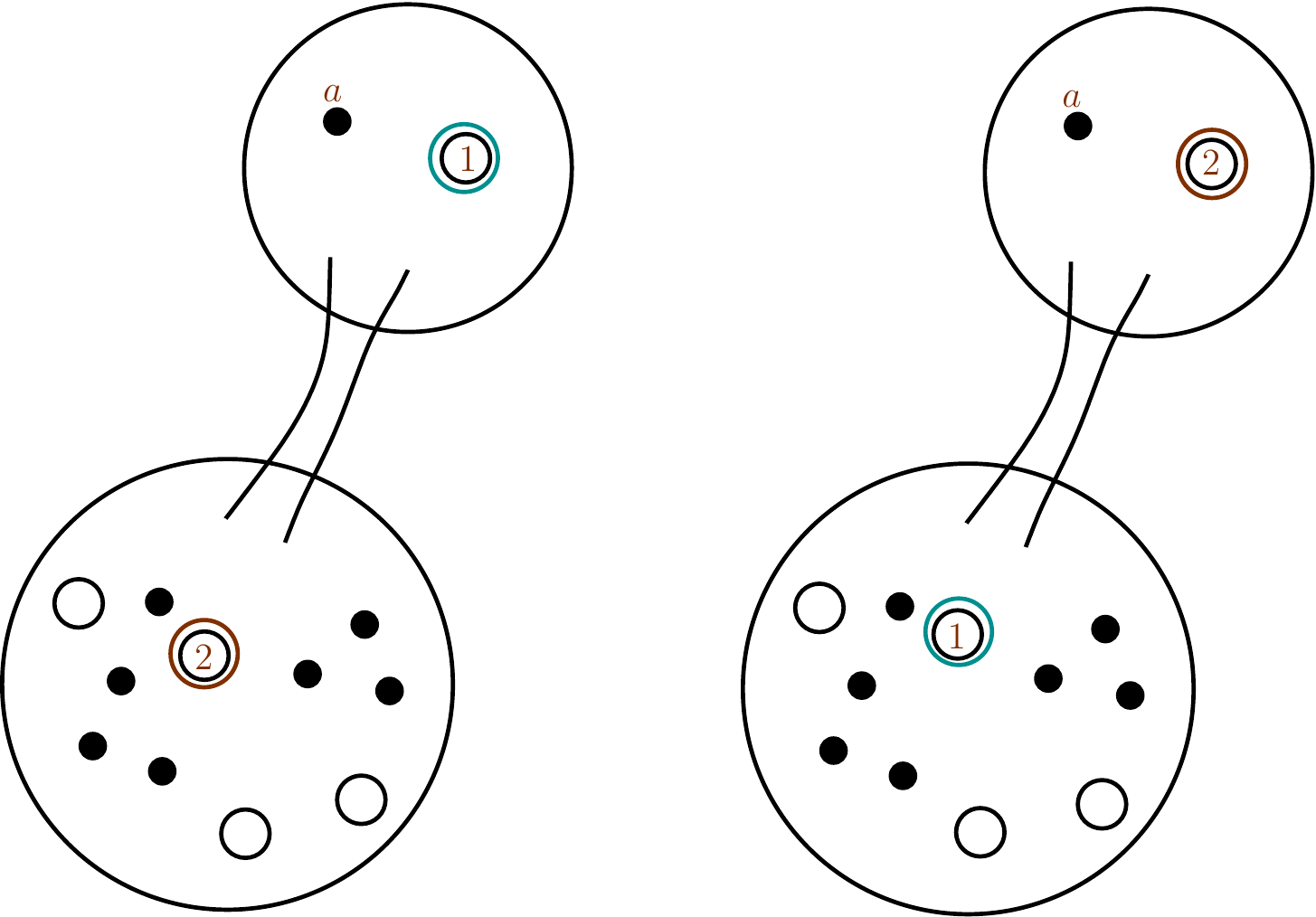}
\end{center}
\caption{Two different pair of pants decompositions corresponding to two different S-duality frames of the field theory.
In the two duality frames the minimal puncture labeled by $a$ sits in a pair-of-pants with a different maximal puncture. 
 The index computed in the two frames should give the same result. }
\end{figure}

As we have already observed, there is nothing special about the puncture labelled by ${\bf x}$. What singled
out ${\bf x}$ in the above calculation is the choice of
a pair-of-pants decomposition where the punctured labelled by ${\bf x}$ belongs to the three-punctured sphere associated to the free hypermultiplet theory.
A different pair-of-pants decomposition would single out a different puncture. By generalized S-duality,
acting with   ${\frak S}_{(r,s)}$ on different punctures must give the same answer:
\be\label{sduop}
{\frak S}_{(r,s)}({\bf x}_k)\,{\cal I}_C({\bf x}_1,\cdots  ,{\bf x}_s)={\frak S}_{(r,s)}({\bf x}_\ell)\,
{\cal I}_C({\bf x}_1, \cdots, {\bf x}_s)\, 
\ee 
for any choice of $k$, $\ell = 1, \dots s$. 
This is  the basic relation that allows to fix the index.

Consider a complete basis of simultaneous eigenfunctions of the difference operators,
\be
{\frak S}_{(r,s)}({\bf x})\, \psi_\lambda({\bf x})={\cal E}_\lambda^{(r,s)}\;\psi_\lambda({\bf x})\, .
\ee
If  the eigenvalues are non-degenerate (as can indeed be  checked to be case),
these functions 
are automatically  orthogonal under the propagator measure, and can be normalized to be orthonormal.
The punchline is now simply stated: this is precisely the Frobenius basis introduced in the previous section
for the TQFT of the index. Indeed, expanding  the index associated to the three-punctured sphere
as
\be
{\cal I} ({\bf x}_1, {\bf x}_2 , {\bf x}_3) = \sum_{\lambda_1, \lambda_2, \lambda_3} C_{\lambda_1 \lambda_2 \lambda_3} \; \psi_{\lambda_1} ( {\bf x}_1)\,
\psi_{\lambda_2} ( {\bf x}_2)\, \psi_{\lambda_3} ( {\bf x}_3) \, ,
\ee
we see from (\ref{sduop}) and the assumption of non-degenerate eigenvalues that the structure constants can be non-vanishing only for
$\lambda_1 = \lambda_2 = \lambda_3$.

The eigenfunctions $\psi_\lambda$ are not known in closed analytic from for general values of the superconformal fugacities $(q, p, t)$,
but there are well-defined algorithms to find them as series expansions (see {\it e.g.}~\cite{Razamat:2013qfa}). Moreover, as we will see in detail
in the following section,  closed analytic  forms are available for special limits of the superconformal fugacities.

\begin{figure}[ht]\label{Tailfig}
\begin{center}
\includegraphics[scale=.7]{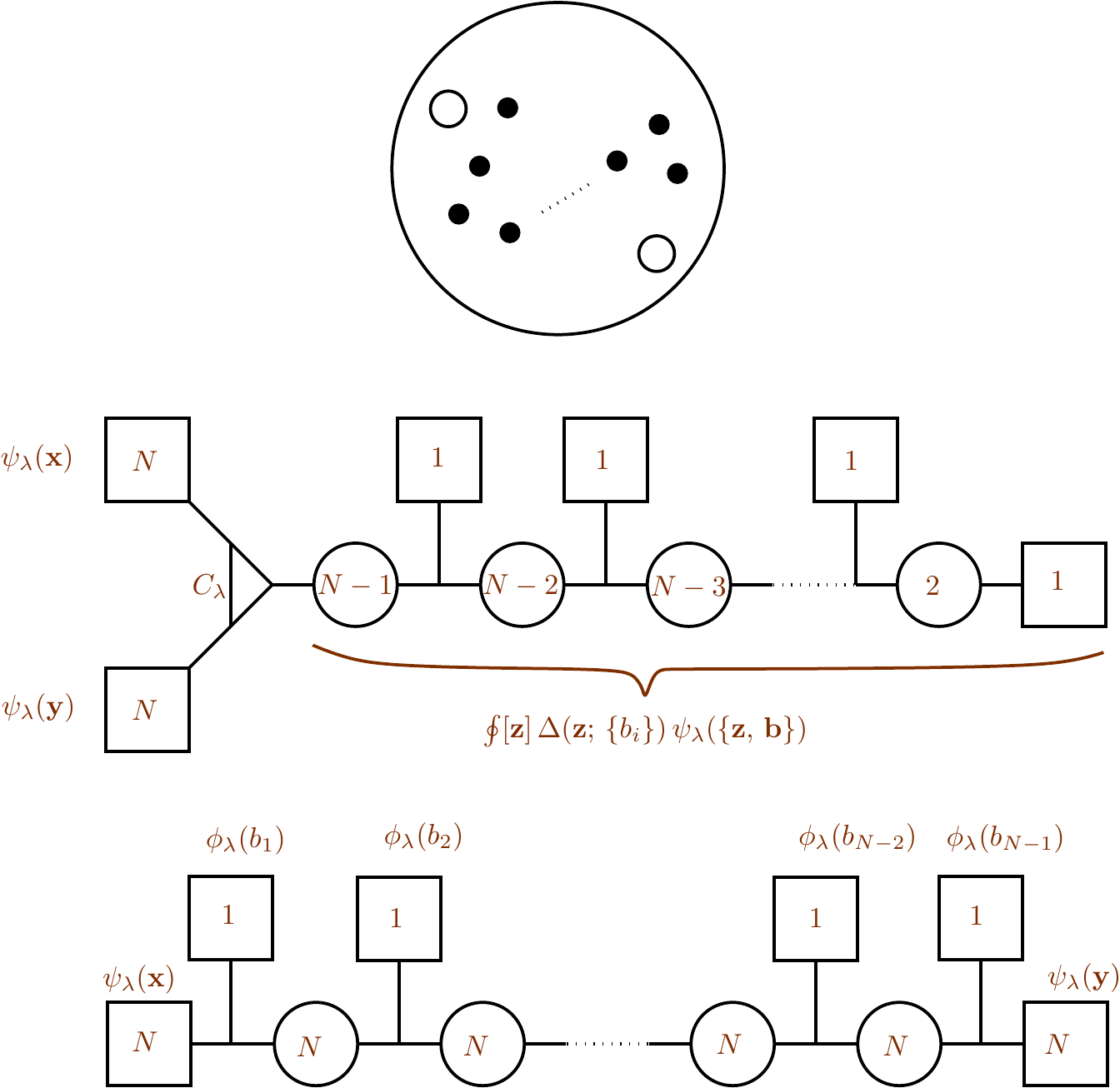}
\end{center}
\caption{One can determine the structure constants $C_\lambda$ of the $A_{n-1}$ theories 
by studying the  theory associated  to a sphere with two maximal and $n-1$ minimal punctures (top picture).
In one duality frame (middle picture) this is given by a $T_n$ theory, involving $C_\lambda$,
 coupled to a ``superconformal tail'' quiver.
In another duality frame (bottom picture) this is given by a linear quiver with  an $SU(n)^{n-2}$ gauge group, where each $SU(n)$ is coupled to $2n$ hypermultiplets.
For $n=3$,  the equivalence of the two frames is the celebrated Argyres-Seiberg duality,
whose consequences for the index  of $T_3$ ($\equiv$ the $E_6$ SCFT) have already  been explored in section \ref{E6}.
  }
\end{figure}

 To complete the computation, it remains  to determine the structure constants $C_\lambda$.
 First,
expanding the index of the free hypermultiplet theory as 
 \be\label{freehypexp}
 {\cal I}_H(a, {\bf x},{\bf y}) = \sum_\lambda \phi_\lambda(a) \, \psi_\lambda ({\bf x}) \,  \psi_\lambda ({\bf y}) \, ,
 \ee
we define the functions $\phi_\lambda(a)$ associated to the minimal puncture.
The functions $\psi_\lambda$ are chosen to be orthonormal under the vector multiplet measure
but functions $\phi_\lambda$ do not have natural normalization properties at this level of the discussion
and their normalization is defined by~\eqref{freehypexp}.\footnote{The same will hold for 
functions $\phi_\lambda^{\Lambda}$ associated to general punctures we will define later in this section.
}
Second, we consider  the theory associated to the sphere with two maximal and $n-1$ minimal punctures.\footnote{We take $n >2$ as the  $n=2$ case is trivial. For $n=2$
there is no distinction between minimal and maximal punctures. 
The basic building block $T_2$ is identified with a free hypermultiplet in the trifundamenal representation of $SU(2)^3$.
The structure constants can then be obtained directly by expanding the free hypermultiplet index.
} 
This theory has two equivalent descriptions, depicted respectively  in  the top and  bottom pictures in Figure~\ref{Tailfig}:  (i) It can be obtained by gluing
to the basic non-Lagrangian building block $T^{[1^n]\, [1^n]\, [1^n]}_n$ a  {\it superconformal tail}~\cite{Gaiotto:2009we},
which is  Lagrangian quiver SCFT with flavor symmetry $SU(n-1) \times U(1)^{n-1}$.   (ii) It 
can be obtained in a completely Lagrangian setup as a linear quiver.
 For the index this implies the following equality:
\be
\sum_\lambda \psi_\lambda({\bf x})\psi_\lambda({\bf y})\prod_{i=1}^{n-1}\phi_\lambda(b_i)=
\sum_\lambda C_\lambda \, \psi_\lambda({\bf x})\psi_\lambda({\bf y}) \int [d{\bf z}]\,\Delta({\bf z};
\, \{b_i\})\,
\psi_\lambda(\{{\bf z},b\})\,, 
\ee where ${\bf z}$ is an $SU(n-1)$ fugacity and an appropriate function of the $b_i$
fixed by matching the $U(1)$ symmetries on the two sides. The function $\Delta({\bf z};
\, \{b_i\})$  can be easily calculated from the superconformal tail.
Since all quantities are known except the structure constants $C_\lambda$, this relation allows to fix them explicitly.
This completes the derivation of the index of class ${\cal S}$ theories of type $A$, with maximal  and minimal punctures.

\

To include punctures of general type $\Lambda$, 
 we need more general  superconformal tails. 
For each $\Lambda$, there exists a minimal  integer $n(\Lambda)$ such that 
the theory associated to 
  one maximal puncture, one puncture of type  $\Lambda$ and $n(\Lambda)$ minimal punctures 
  can be described by   a  Lagrangian quiver gauge theory~\cite{Gaiotto:2009we}. 
  This can in fact be viewed as a {\it definition} of the puncture of type $\Lambda$.
 By equating the abstract definition of the index of such a theory, namely
   \be
\sum_\lambda \psi_\lambda({\bf x})\, \phi_{\lambda}^{\Lambda}({\bf y}_\Lambda)
\, \prod_{i=1}^{n(\Lambda)}\phi_\lambda(b_i) \, ,
\ee 
with the explicit integral expression of the same index given by Lagrangian quiver description
we can determine the factor $\phi_{\lambda}^{\Lambda}({\bf y}_\Lambda)$ associated to the puncture of type $\Lambda$.

\

In summary, we have described an algorithm that determines the superconformal index
for all theories of class ${\cal S}$ with regular punctures. The index takes an elegant general form
in terms of structure constants $C_\lambda (p, q, t)$ and of
``wavefunctions'' $\{\phi^{\Lambda_i}_\lambda({\bf y}_{\Lambda_i}; p, q, t)\}$ associated
to the punctures,\footnote{Comparing with (\ref{Ipsi}),
we have reabsorbed some factors of $C_\lambda$ into  wavefunctions, by setting a new normalization
for the wave function of the maximal puncture,
$\phi_\lambda^{[1^n]} \colonequals  C_\lambda \psi_\lambda$.}
\be\label{finindex}
{\cal I}=\sum_\lambda C_\lambda^{2{g}-2}\,\prod_{j=1}^{s}\phi^{\Lambda_i}_\lambda({\bf y}_{\Lambda_i})\, ,
\ee
where the sum is over the set of finite-dimensional irreps of $\gf = \suf(n)$.

A caveat is in order. Not every possible choice of   Riemann surface decorated by a choice
of $\{ {\Lambda}_i \}$ at the punctures
corresponds to a physical SCFT.  
An indication that a choice of decorated surface may be unphysical is if the sum  in (\ref{finindex}) diverges,
which happens when the  flavor  symmetry is ``too small''. There are subtle borderline
cases where the sum diverges, but the theory is perfectly physical -- this can happen
when the theory has additional ``accidental'' flavor symmetries not associated to punctures. An 
example of such a theory is the rank two $E_6$ SCFT. These cases have to be treated with more care~\cite{Gaiotto:2012uq}.

We will discuss how to calculate explicit expressions for the wavefunctions and structure constants in the next section.
In the rest of this section we  offer two viewpoints that illuminate the structure of the result,
the first related to Higgsing and the second  to dimensional reduction.

\

\subsection{Higgsing: reduced punctures and surface defects}\label{surfacesec}

The index is a meromorphic function of flavor and superconformal fugacities,
with a rich structure of poles. 
A large class of these poles  has a nice physical interpretation \cite{Gaiotto:2012xa}.

Consider a schematic version of the index,
 \be
{\cal I}(a,b)=\Tr(-1)^F\, a^f\,t^{R}\,,
\ee where $f$ and $R$ are two conserved charges. 
Let us assume that ${\cal I}$ has a pole in fugacity $a$,
\be \label{toypole}
{\cal I}=\frac{\widetilde {\cal I}(a,\, t)}{1-a^{f_{\cal O}}\, t^{R_{\cal O}}}\,.
\ee
It is  natural to associate the pole to a bosonic
operator ${\cal O}$, with charges $f=f_{\cal O}$ and  $R=R_{\cal O}$,
such that an infinite tower of composites  of the form ${\cal O}^n$ contribute to the index.
In the simplest case, ${\cal O}$ is the generator of a  ring spanned by $\{ {\cal O}^n \}$, and the pole
appears by resumming the geometric sum,
\be
1 + a^{f_{\cal O}}\, t^{R_{\cal O}} + (a^{f_{\cal O}}\, t^{R_{\cal O}})^2 + \dots
\ee
In more complicated cases, there can be several generators obeying non-trivial relations, which are encoded
in the numerator of (\ref{toypole}).
 The residue  at  $a^{f_{\cal O}}\, t^{R_{\cal O}} =1$
  is given by
$  \widetilde {\cal I} (
t^{-R_{\cal O}/f_{\cal O}}
,\, t)$, which can be interpreted as
\be \label{redefine}
\Tr'(-1)^F\,\,t^{\bar R}\, ,\quad  \bar R \colonequals  R-\frac{R_{\cal O}}{f_{\cal O}}\, f\, ,
\ee 
where the prime on the trace indicates that we are omitting the infinite set of states with $\bar R= 0$, which
are of course the states responsible for the pole in the first place.
The shifted charge $\bar R$ is the linear combination of charges preserved in a background where ${\cal O}$ has acquired a non-zero vacuum expectation value (vev). In  a path integral representation of the index as the $\S^3 \times \S^1$ partition function,
the divergence at  $a^{f_{\cal O}}\, t^{R_{\cal O}} =1$ arises from the integration over a bosonic zero mode,
which heuristically we identify with  $\langle {\cal O} \rangle$. Following this intuition, 
we expect the {\it residue}  to be controlled by the behavior of theory ``at infinity'' in the moduli space 
parametrized by $\langle {\cal O} \rangle$,
that is, by the properties of  the  IR theory  reached at the endpoint of the 
 the RG flow triggered by giving   ${\cal O}$ a vev.  We interpret  $\widetilde {\cal I}$ as the index of this IR fixed point.

\

\noindent {\it Reducing punctures}

\

\noindent As a first application of  these ideas, let us obtain more directly  the index in the presence
of punctures of general type, taking as starting point the index with  maximal punctures.
The idea is that the theory with a partially-closed puncture can be obtained from the theory
with a full puncture by partially higgsing the full $\suf(n)$ flavor symmetry, and flowing to the IR.\footnote{The equivalence
between the realization of general punctures by superconformal tails (as sketched in the previous subsection)
and the higgsing procedure  that we are about to implement is explained in section 12.5 of  \cite{Tachikawa:2013kta}.}
 The role of the operator ${\cal O}$ that featured the above general discussion is played
 by the {\it moment map} operator $\mu$. The moment map is the superconformal primary
 of the supermultiplet that contains the flavor symmetry current, and thus transforms in the adjoint representation of $\suf(n)$.\footnote{
 The moment map is also an $SU(2)_R$ triplet and $U(1)_r$ singlet. We consider the highest $SU(2)_R$ weight (which has $R=1$),
 since it is the component that contributes to the index.}
 Given an embedding $\Lambda : \suf(2) \to \suf (n)$, we choose the vev of $\mu$ to be
\be
 \langle \mu \rangle = \Lambda (t^-) \in {\rm adj}_{\suf(n)} \, ,
  \ee
where $t^-$ is the lowest weight of $\suf(2)$. The flavor symmetry is broken down to the centralizer
of $\Lambda$ in $\suf(n)$, which we call $\gf_\Lambda$. We expect to find poles
in the wavefunction  $\phi_\lambda ( {\bf a} )$  in correspondence to each component of $\mu$ that receives a vev.
Extracting the residues with respect to such poles should give the 
wave function  $\phi_\lambda^{\Lambda} ( {\bf x}_\Lambda)$ associated to the reduced puncture.
More precisely, the symmetry breaking also generates Goldstone modes that give a decoupled free sector, and we should remove
 their contribution if we are interested in the interacting IR SCFT. Finally we should remember
 to redefine charges, following the general principle outlined in (\ref{redefine}).  In our case, the vev for $\mu$ breaks
 the $SU(2)_R$ symmetry, however a linear combination $\bar R$ of the original $R$ Cartan generator and of flavor Cartan generators
 is preserved; we expect this symmetry to enhance in the IR to the full non-abelian $SU(2)_{\bar R}$ of the interacting fixed point.\footnote{
It might be that the vev actually preserves the diagonal subgroup of the UV $su(2)$ R-symmetry and
some $su(2)$  subgroup of the flavor symmetry. In such a case there is no need for the IR enhancement of 
the R-symmetry. We thank C.~Beem, D.~Gaiotto, and A.~Neitzke for pointing this out to us. 
}
 All in all, we have the prescription
\be \label{waveprescription}
G^\Lambda ( {\bf a}_{\Lambda}  )\, \phi^{\Lambda}_\lambda({\bf a}_{\Lambda}) =  {\rm Res}_{ {\bf a} \to  \mathrm{fug}_{{\Lambda}}(\mathbf{a}_{{\Lambda}},t)  } \, 
\phi_\lambda ( {\bf a} ) \, ,
\ee
where the prefactor  $G^\Lambda$, which is easily computable, accounts for the contribution  to the index of the Goldstone bosons induced
by the symmetry breaking. The fugacity replacement ${\bf a } \to  \mathrm{fug}_{{\Lambda}}(\mathbf{a}_{{\Lambda}},t)$
can be obtained with a little representation theory.
Any representation $\mathfrak{R}$ of $\mathfrak{g}= \suf(n)$  decomposes as
\begin{equation}\label{generaldecomposition}
\mathfrak{R} = \bigoplus_j \mathcal{R}_j^{(\mathfrak{R})} \otimes V_j\;,
\end{equation}
where $\mathcal{R}_j^{(\mathfrak{R})} $ is some (generally reducible) representation of $\mathfrak{g}_{\Lambda}$ and $V_j$ the spin $j$ representation
of $\suf(2)$. Then
$\mathrm{fug}_{{\Lambda}}(\mathbf{a}_{{\Lambda}},t)$ is  the  solution for  $\mathbf{a}$ in the character decomposition equation,\footnote{The solution is unique up to  the action of  the Weyl group.}
\begin{equation}\label{definefug}
\chi_{ \mathfrak{f} }^{\mathfrak{g}}(\mathbf{a}) =  \sum_j \ \chi_{\mathcal{R}_j^{(  \mathfrak{f} )}}^{\mathfrak{g}_{{\Lambda}}}(\mathbf{\mathbf{a}}_{{\Lambda}}) \, \chi_{V_j}^{\mathfrak{su}(2)}(t^{\frac{1}{2}}\, ,t ^{-\frac{1}{2}})\; .
\end{equation}
One can check that (\ref{waveprescription}) reproduces the wavefunctions obtained using 
superconformal tails by the method
 outlined in the previous subsection. Let us give a couple of simple examples.

Taking $\gf = \suf(2)$ and $\Lambda: \suf(2) \to \suf(2)$ the principal embedding, which in this  case is just the identity map,
the centralizer is of course trivial and (\ref{definefug}) reads
\be
a + a^{-1} = t^{\half} + t^{-\half} \,, 
\ee
 which has  the two solutions $a = t^ {\half}\, , t^{-\half}\,$, related by the action
 of  the Weyl group $a \leftrightarrow a^{-1}$.  Since we are interested
 in the vev of the  $\suf(2)$ lowest weight  $\mu^{-}$ of the moment map, whose contribution to the index is $a^{-2} t$,
 we should pick $a = t^{\half}$; the other solution   $t^{-\half}$ would be associated to the $\suf(2)$ highest weight  $\mu^{+}$.
 The lesson (which generalizes) is that if we are interested in giving a vev to specific operator,  we should 
 fix a representative of the Weyl orbit.
 Extracting the residue at  $a^2 t^{-1} =1$ will give the index of the IR theory at the end of the RG flow triggered by $\langle \mu^- \rangle$,
{\it times} the contribution from the free Goldstone bosons. In this case, the Goldstone bosons
consist of a free hypermultiplet in the fundamental of the flavor $\suf(2)$. Both the flavor and $R$ symmetry are broken
by the vev, but the combination $\bar R = R +f/2$ is preserved. 
  Under the new $SU(2)_{\bar R}$, the scalars of the free hypermultiplet transform as ${\bf 3} + {\bf 1}$,
 with the singlet corresponding to the states responsible for the divergence. Extracting the pole
 is precisely equivalent to omitting this singlet states. Setting $a=t^\half$ in (\ref{spindices})
  we see that
 under this new charge assignment the non-singlet states of the free hypermultiplet  give a contribution
 to the index exactly equal to  the inverse of the index of a free $U(1)$ vector multiplet,
 so  the Goldstone boson factor in (\ref{waveprescription}) is $G ={\cal I}_V^{-1}$.
  All in all, we have derived from general
 principles the following prescription to close an $\suf(2)$ puncture,
\be \label{closingsu2}
 {\cal I}_V\,  {\rm Res}_{a^{-2}  t \to 1 } \,\phi^{[1^2]}_\lambda (a ) =  \phi_\lambda^{[2]} \equiv 1 \, .
 \ee
In the last equality we have just  reminded ourselves that the wavefunction of a  fully closed puncture is identically equal to one.
One can check (\ref{closingsu2})
using  the expression for $\phi^{[1^2]}_\lambda$ derived by the methods of the previous subsection.

A sightly more involved example is $\gf = \suf(3)$
and $\Lambda$ the subregular embedding, corresponding to the partition $[2, 1]$. The centralizer is
 $\gf_\Lambda = \uf (1)$. If $a_1$, $a_2$, $a_3$ with $a_1 a_2 a_3 = 1$ are the $\suf(3)$ fugacities, and $b$ the $\uf(1)$ fugacity,
(\ref{definefug}) takes the form
\be
a_1 + a_2 + a_3 = b \, ( t^{\half} + t^{-\half} ) + b^{-2}   \, .
\ee
The only solution (up to the action of the Weyl group, which permutes the $a_i$)  is
$a_1 = t^{\half} b$,  $a_2 = t^{-\half} b$, $a_3 = b^{-2}$.  
 Extracting the residue and removing the contribution of the Goldstone bosons
  accomplishes the reduction of the full puncture to the minimal puncture.

\

\noindent {\it Surface defects}

\

\noindent  Next, we would like to  interpret in a similar light  the poles (\ref{apoles})
that played such a crucial role in the previous subsection. Recall the basic setup: we ``glued''
the  bifundamental hypermultiplet theory $T_n^{[1^n] \, [1^n] \, [ n-1, 1]}$
 to a general theory $T[{\cal C}]$, connecting
 a maximal puncture of one theory with a maximal puncture of the other theory  by gauging the diagonal $SU(n)$ symmetry.
We then extracted residues with respect to
the fugacity $a$  for the $U(1)$ global symmetry of the hypermultiplet.  This is the
$U(1)$ baryon symmetry, under which the complex scalars $q$ and $\tilde q$ have charge $-1$ and $+1$ respectively.
It is then clear that  the operator associated to the
simplest pole, at $a^{-n} t^{\frac{n}{2}} =1$, is 
 the baryon operator $B = {\rm det}\; q$. Giving a vev to $B$  
higgses the $SU(n)$ gauge group, triggering an RG flow whose IR endpoint 
is the original  theory $T[{\cal C}]$ and a collection of decoupled free fields~\cite{Gaiotto:2012xa}. 
This explains~\eqref{basres}.\footnote{For $n=2$, the $U(1)$ baryon symmetry enhances to $SU(2)$,
$B \equiv \mu^-$ (the lowest weight component of the moment map), 
and  \eqref{basres} is precisely equivalent to (\ref{closingsu2}).}

\

By the same logic,  the poles at $ p^{r} q^{s}   t^{\frac{n}{2}} a^{- n}  = 1$ 
are naturally associated to holomorphic derivatives of the baryon operator  in the $12$ and $34$ planes,  
 $\partial^r_{12}\,\partial^s_{34}\,\det q$. 
We expect the residue at these poles to describe the IR physics of the flow triggered
by  a spacetime-dependent vev of the form
 $\langle B  \rangle \sim z^r w^s$.  Consider first the $r=0$, $s \neq 0$ case.  Away from the $w=0$ plane,
 the endpoint of the flow is still $T[{\cal C}]$. However some
 extra degrees of freedom survive at $w=0$, which we interpret as a {\it surface defect} for $T[{\cal C}]$
  extended in the 12 plane.  Similarly, the endpoint of the flow with $r\neq 0$, $s =0$ is $T[{\cal C}]$ 
  decorated with an extra surface defect extended in the 34 plane. In the general case with $r s \neq 0$ both type
  of defects will be present. 
 In the $\S^3 \times \S^1$ geometry,
 these surface defects  fill the ``temporal'' $\S^1$ and the two maximal circles inside the $\S^3$ fixed by the $j_{12}$ and $j_{34}$ rotations,
 respectively.  This proposal has been checked~\cite{Gadde:2013dda} in a set of examples
 where $T[{\cal C}]$ admits a Lagrangian description, and surface defects
 can be added by coupling the $4d$ SCFT to a $(2, 2)$ sigma model; the index
 can then be independently evaluated by localization techniques, confirming the prescription
  that we have just outlined.

\

In summary, we have found a  physical interpretation for the  difference operators ${\frak S}_{(r,s)}$:
their action on the index of $T[{\cal C}]$
yields  the index of the same theory decorated by  some extra surface
 defect~\cite{Gaiotto:2012xa}. 
 Since the difference operators act ``locally'' on the generalized
 quiver, we should associate them to special punctures of the UV curve. This agrees with the M-theory picture,
 where the surface defects correspond to M2 branes localized on the UV curve.
 Acting with a difference operator on a given flavor fugacity corresponds pictorially to colliding
 the special puncture with a flavor puncture.
 The location of the special punctures on the UV curve
 is immaterial, so collision of the same special puncture with different flavor punctures
 is bound to give the same result -- which is a restatement of (\ref{sduop}).

\

This description of surface defects bears a striking kinship
with the analogous picture that arises in the AGT correspondence ~\cite{Alday:2009aq,
Alday:2009fs}. The introduction of surface defects in the $\S^4$ partition function
is accomplished by the insertion of special, semi-degenerate operators
in the Toda  CFT  correlator defined on the UV curve. 
 These operators are the key to the solution of Liouville
theory by the conformal bootstrap~\cite{Teschner:1995yf}: considering their fusion with normalizable vertex
operators one can derive functional equations that admit a unique solution. Similarly,
we have special punctures in our 2d TQFT that insert 
surface defects in the $\S^3 \times \S^1$ partition function. Their fusion
with  ordinary  flavor punctures  leads to  the topological  bootstrap equations (\ref{closingsu2}),
which uniquely fix the superconformal index.

\begin{figure}[ht]\label{Differences}
\begin{center}
\includegraphics[scale=.5]{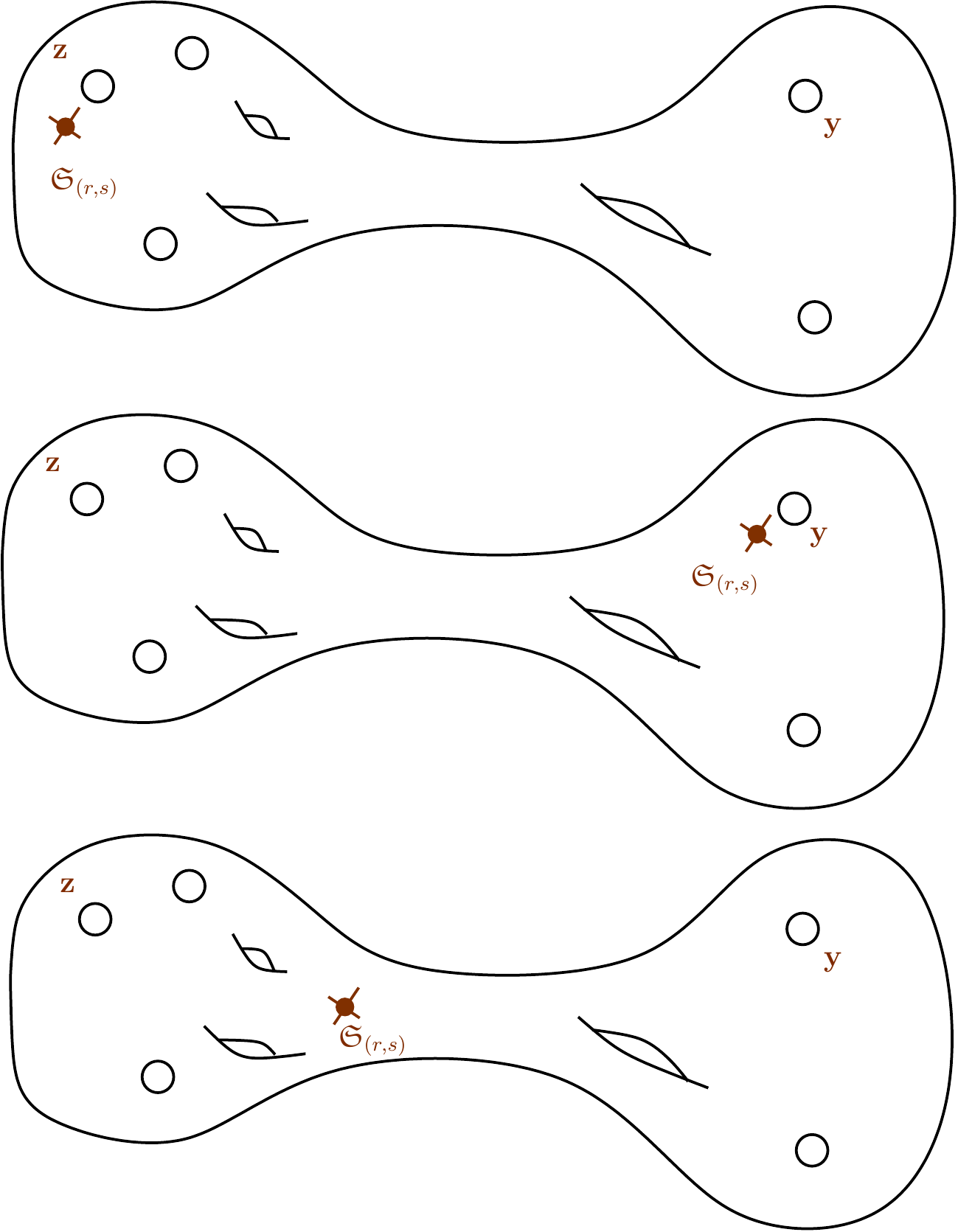}
\end{center}
\caption{The difference operators  ${\frak S}_{(r,s)}$, which
compute residues and introduce surface defects, can be visualized as special punctures on the UV curve.
The action of  ${\frak S}_{(r,s)}$ on a flavor fugacity is interpreted as the collision
of the special puncture 
 with a flavor puncture.  We can  act on different punctures and obtain the same result for the index (top and middle pictures).
 We can also define the action of  ${\frak S}_{(r,s)}$ on a long tube   (bottom picture), 
  by cutting open a cylinder, acting on one of the open punctures and gluing the surface back.
S-duality guarantees that this is a well-defined procedure.
In this way we can introduce the special punctures ${\frak S}_{(r,s)}$ on a UV curve  with no flavor punctures at all.
}
\end{figure}

\subsection{Reduction to 3d}\label{threesec}

The index of theories of class ${\cal S}$ has a very definite structure~\eqref{finindex}. This structure
is natural since it is a manifestation of the $2d$ TQFT nature of the index of the theories at hand
as was anticipated in section 4.2. It is however an important question to understand better the physical 
meaning of the different ingredients entering~\eqref{finindex}. For example, we would like to gain
more insight  into the physical significance of the eigenfunctions $\phi^{\Lambda_i}_\lambda({\bf y}_{\Lambda_i})$ and the eigenvalues ${\cal E}_\lambda^{(r,s)}$. 
Let us  consider here a very informative $3d$ interpretation  of~\eqref{finindex}.\footnote{
A $6d$ physical interpretation of this equation can be also entertained~\cite{GRR6d} but we will not discuss it
in this review.
}

We can consider theories of class ${\cal S}$ on ${\cal M}_3\times \S^1$ with ${\cal M}_3$ some three dimensional manifold. Upon reduction on the $\S^1$ we obtain a $3d$ theory on ${\cal M}_3$. 
The ${\cal N}=2$ class ${\cal S}$ theories admitting a known description in terms of a Lagrangian 
upon dimensional reduction on $\S^1$ are described in terms of the same field content and same gauge 
and superpotential interaction as the $4d$ parent theory. The $3d$ Lagrangians however are not conformal
and the theories flow in general to an interacting ${\cal N}=4$ $3d$ SCFT in the IR. The $4d$ conformal  S-dualities imply IR (Seiberg-like) dualities of the $3d$ models. Thus the complex moduli of the Riemann 
surface defining the model in $4d$ do not translate to physical parameters in $3d$: the topology of the surface and the information at the punctures alone are sufficient to completely specify the $3d$ model.
An extremely interesting fact about the  class ${\cal S}$ theories in $3d$ is that they possess yet another dual description. All theories of class ${\cal S}$ reduced to $3d$, with and without known Lagrangian description in $4d$, have a mirror description in $3d$ in terms 
of a {\it star-shaped} quiver theory~\cite{Benini:2010uu}. 

\begin{figure}[ht]\label{mirr}
\begin{center}
\includegraphics[scale=.8]{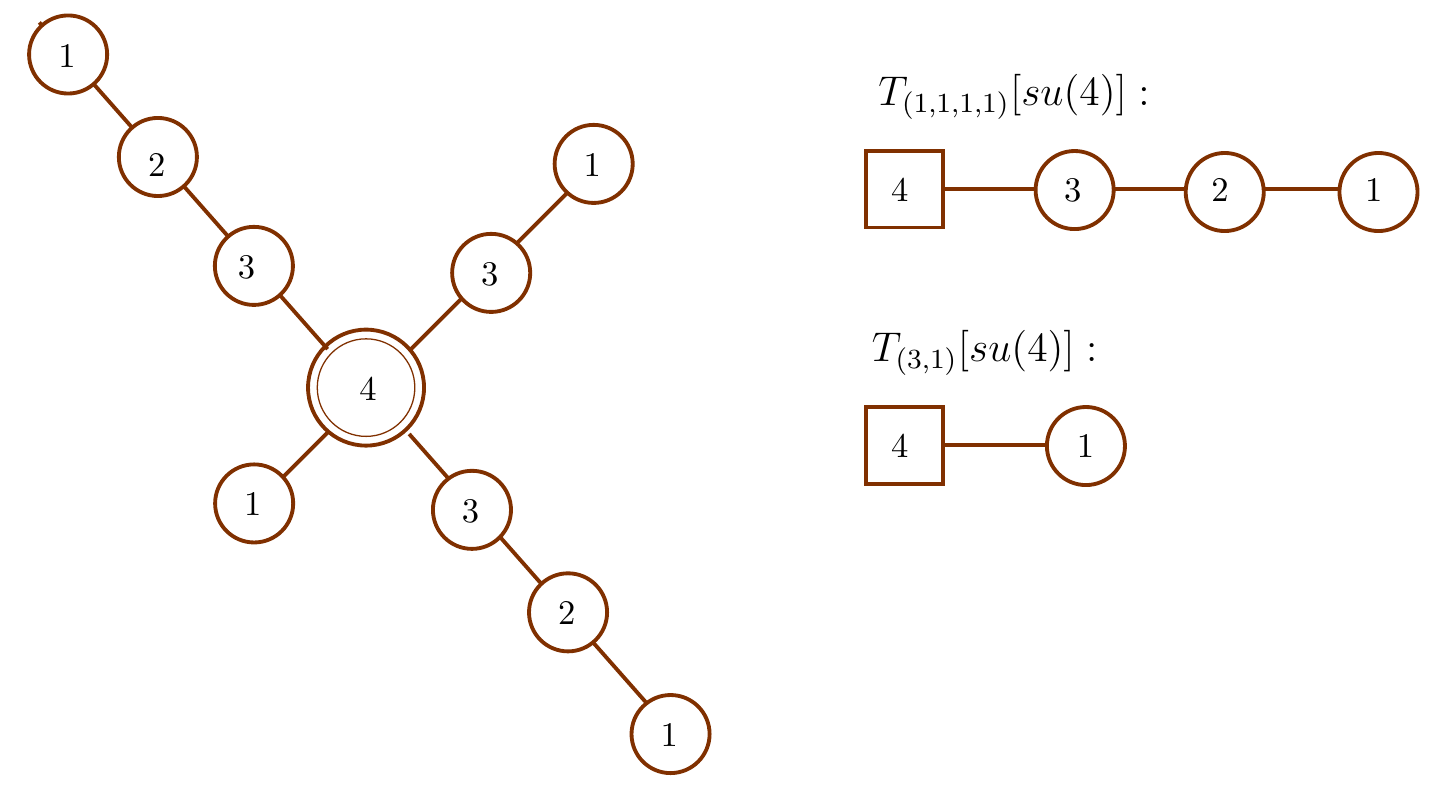}
\end{center}
\caption{On the left we have an example of a star-shaped quiver mirror  of the $A_3$ theory corresponding to a sphere with four punctures, two of which are maximal and one is minimal.
On the right the quiver theories for ${\frak T}_{(1,1,1,1)}[\suf(4)]$ corresponding to the maximal puncture and
${\frak T}_{(3,1)}[\suf(4)]$ corresponding to minimal puncture are depicted.}
\end{figure}

This mirror symmetry states that a theory corresponding to  a Riemann surface with genus $g$ and $s$ punctures of types $\Lambda_i$ is dual to a quiver theory coupling $s$ 
linear quivers ${\frak T}_{\Lambda_i}[{\frak g}]$~\cite{Gaiotto:2008ak} associated to Lie  algebra ${\frak g}=\suf(N)$  by gauging  the common ${\frak g}$ with an addition of $g$ ${\cal N}=4$ adjoint hypermultiplet, see figure~\ref{mirr} for an example.

The dimensional reduction on $\S^1$ can be performed at the level of the index. Here ${\cal M}_3$ is 
$\S^3$ and upon reduction of the index on $\S^1$ we obtain the partition function of the dimensionally reduced theory on a squashed sphere $\S^3_b$~\cite{Dolan:2011rp,Gadde:2011ia,Imamura:2011uw} (see also~\cite{Aharony:2013dha}). The reduction is done by first parametrizing the fugacities as
\be
 t=e^{2\pi i  r_1 (\gamma + \frac{i}{2 r_3}(b+b^{-1}))},  \;\;\; z=e^{2\pi i  r_1 \sigma}, \;\;\; p=e^{2\pi b  r_1/r_3}, \;\;\; q=e^{2\pi b^{-1}   r_1/r_3} \,,
\ee where $r_1$ is the radius of $\S^1$ and $r_3$ is the radius of $\S^3$. Then the radius of $\S^1$, $r_1$, is sent to zero. The parameter $b$ is the squashing parameter of the sphere.

We have defined the functions $\phi^{\Lambda_i}_\lambda({\bf y}_{\Lambda_i})$ as eigenfunctions
of difference operators ${\frak S}_{(r,s)}$ and argued that this operators have a physical 
interpretation of introducing linked surface defects to the index computation. The surface defects
corresponding to ${\frak S}_{(0,s)}$ and ${\frak S}_{(r,0)}$ span the $\S^1$ and one of the two equators of $\S^3$. Upon reduction on the $\S^1$ these become line defects sitting on one of the two equators of
$\S^3_b$. When sending $r_1\to 0$ the difference operators have very simple limit. For example
in the $A_1$ case we have\footnote{This operator is called the Macdonald operator in math literature and
we will shortly encounter a different incarnation of it in $4d$ index context.}
\bea\label{3dop}
&&{\frak S}_{(0,1)}\cdot f(z)\qquad\to\qquad {\cal T}(\sigma)\cdot\hat f(\sigma)=\\
&&\,\qquad
=\frac{\sinh\pi b\left(\frac{i(b-b^{-1})}2-\gamma+2\sigma\right)}{\sinh2\pi b\sigma} \hat f(\sigma+\frac{i b}{2})+
\frac{\sinh\pi b\left(\frac{i(b-b^{-1})}2-\gamma-2\sigma\right)}{\sinh-2\pi b\sigma}\hat f(\sigma-\frac{i b}{2})\,,\nonumber
\eea where $\hat f(\sigma)=\lim_{r_1\to 0}f(e^{2\pi i r_1\sigma})$.
Interestingly a set of eigenfunctions of this operator is given by the $\S^3_b$ partition
functions of the ${\frak T}[\suf(2)]$ theory. The ${\frak T}[\suf(N)]$ theory has global $\suf(N)_H\times \suf(N)_C$ symmetry
with the $\suf(N)_H$ acting on the Higgs branch and $\suf(N)_C$ acting on the Coulomb branch. Turning on real mass parameters, ${\bf \sigma}_H$ and ${\bf \sigma}_C$, for the two symmetries the $\S^3_b$ partition function of 
${\frak T}[\suf(N)]$ can be denoted by $\phi^{({\gamma,b})}({\bf \sigma}_H|{\bf \sigma}_C)$ and we have the property 
\be
 {\cal T}({\bf \sigma}_C)
\cdot  \phi^{({\gamma,b})}({\bf \sigma}_H|{\bf \sigma}_C)= {\cal W}({\bf \sigma}_H)\,
\phi^{({\gamma,b})}({\bf \sigma}_H|{\bf \sigma}_C)\,.
\ee The eigenvalue ${\cal W}({\bf \sigma}_H)$ is the expectation value of the Wilson
loop for the $\suf(N)_H$ global symmetry. This eigenvalue property of the 
partition function thus suggests the physical interpretation
that the line defect for the gauge symmetry of ${\frak T}[\suf(N)]$ is equivalent to 
 a Wilson line for the global $\suf(N)_H$ symmetry. This fact is not surprising since 
the ${\frak T}[\suf(N)]$ theories make their appearance as models living on S-duality domain
wall separating two S-dual ${\cal N}=4$ $SU(N)$ SYM  theories. Since under $4d$ S-duality
defect  ('t Hooft) line operators map to Wilson operators our $3d$ eigenvalue statement
is natural.

Further, the $\S^3_b$ partition function of a star shaped quiver mirror  dual say to the $A_1$ theory 
with genus $g$ and $s$ punctures has the following form,
\be\label{sqpart}
Z_{g,s}(\{\sigma^{(i)}_C\}_{i=1}^s)=
\int d\sigma_H Z_V(\sigma_H)\; Z_{\cal H}(\sigma_H)^g\; \prod_{i=1}^s\phi^{(\gamma,b)}(\sigma_H|\sigma^{(i)}_C)\,.
\ee Here $Z_{\cal H}(\sigma_H)$ is the contribution of an ${\cal N}=4$ $3d$ 
adjoint hypermultiplet and  $Z_{V}(\sigma_H)$ is the contribution of the vector.
Note the striking structural similarity between~\eqref{finindex} and~\eqref{sqpart}.
This is not a coincidence~\cite{Nishioka:2011dq}. One can argue that indeed in the $r_1\to0$ limit
the eigenfunctions $\phi^{\Lambda}_\lambda({\bf y}_{\Lambda})$ reduce to the $\S^3_b$ partition functions of ${\frak T}_\Lambda[\suf(N)]$. The discrete labels of the eigenfunctions, $\lambda_i$, become (linear combinations of) the real masses of the symmetry rotating the
Coulomb branch of ${\frak T}_\Lambda[\suf(N)]$, $\sigma^{(i)}_C$: roughly, taking the $r_1\to0$ limit we 
should also concentrate on large representations and keep $r_1\, \lambda_i$ fixed.

Let us summarize the $3d$ interpretation of the eigenfunctions,
\begin{itemize}
\item The difference operators introduce line defects.
\item The eigenfunctions are $\S^3_b$ partition functions of ${\frak T}_\Lambda[\suf(N)]$.
\item The eigenvalues are expectation values of Wilson loops.
\item The existence of the eigenvalue equation follows from $4d$ S-duality through the statement 
that Wilson and 't Hooft lines are S-dual to each other.\footnote{When writing this equation as a difference operator anihilating  the partition function, it gives rise actually to the difference 
operator  anihilating holomorphic blocks of the $3d$ partition function~\cite{Beem:2012mb}.
}
\end{itemize} In particular the fact that the index of theories of class ${\cal S}$ in $4d$ can be written in the form~\eqref{finindex}  is a $4d$ manifestation of the fact that the dimensionally reduced theories admit a mirror description. {\it That is the index written as~\eqref{finindex} is a $4d$ precursor 
of the $3d$ mirror symmetry.} The interested reader might consult~\cite{Razamat:2014pta}
 for more thorough 
discussion of these issues.

Finally let us also mention that the $3d$ eigenfunctions, $\S^3_b$ partition functions of ${\frak T}_\Lambda[\suf(N)]$, provide a connection between the $4d$ index and the $4d$ $\S^4$ partition 
functions of theories of class ${\cal S}$.  As we mentioned ${\frak T}_\Lambda[\suf(N)]$ models are obtained 
by considering ${\cal N}=4$ $4d$ theories with a duality domain wall. The kernel which implements
the insertion of such duality wall in the $\S^4$ partition function computation is precisely the $\S^3_b$
partition function of ${\frak T}_\Lambda[\suf(N)]$~\cite{Hosomichi:2010vh}. In particular the difference operator we obtained 
by reduction to $3d$ are the same difference operators introducing line defects into Liouville-Toda/$\S^4$ (AGT correspondence~\cite{Alday:2009aq}) computations~\cite{Gaiotto:2012xa} (see also~\cite{Bullimore:2014nla}).

\section{Integrable models and  limits of the index}\label{secLimits}

The discussion of the previous section reduces the physical problem of determining the superconformal
index of class ${\cal S}$ theories to the mathematical problem of finding 
a complete set of orthonormal eigenfunctions of  the difference operators ${\frak S}_{(r,s)}({\bf x})$. Remarkably, these
operators are closely related to the Hamiltonians that define a well-known class of integrable models, the
elliptic relativistic Ruijsenaars-Schneider (RS) models, {\it aka} relativistic elliptic Calogero-Moser-Sutherland models.

The  operator~\eqref{basicOp}, ${\frak S}_{(0,1)}({\bf x})$, is related to the basic  RS Hamiltonian ${\cal H}_1(t,q;p)$ 
by a similarity transformation,
\be\label{conj}
{\cal H}_1(t,q;p)=\frac{\theta(q^{-1};p)}{\theta(t;p)}\;\frac{1}{\prod_{i\neq j}\Gamma(t\,z_i/z_j;p,\,q)}\;
{\frak S}_{(0,1)}({\mathbf z})\;
\prod_{i\neq j}\Gamma(t\,z_i/z_j;p,\,q)\,.
\ee 
Under the same similarity transformation, the propagator measure  in the $A_{n-1}$ case becomes
\be\label{meas2}
\frac{1}{n!}\oint \prod_{i=1}^{n-1}\frac{dz_i}{2\pi i z_i}
\prod_{i\neq j}\frac{\Gamma(t\,z_i/z_j;p,\,q)}{\Gamma(z_i/z_j;p,\,q)}\,\cdots\,.
\ee
Higher operators, ${\cal H}_\ell$, can be constructed as polynomials in ${\frak S}_{(0,s)}$. 
One can think of the $n-1$ independent ${\cal H}_\ell$ operators as associated to antisymmetric representations of $SU(n)$, whereas ${\frak S}_{(0,s)}$ are associated to symmetric representations. 
Then by exploiting group theory and the fact that the fundamental representation can be trivially thought as either symmetric or antisymmetric one can translate between ${\cal H}_\ell$ and
 ${\frak S}_{(0,s)}$ (see for example~\cite{Bullimore:2014nla}). 

 The parameters $p$, $q$, and $t$  appear in the Hamiltonian ${\cal H}_1(t,q;p)$ on different footing:
(i) the parameter $t$ plays a role of coupling constant, (ii) $q$ is the shift parameter of the difference operator and can be understood as an exponent of the ``speed of light'' parameter of the relativistic integrable system, (iii) the integrable model is associated to an elliptic curve parametrized by $p$.
Given an eigenfunction of ${\cal H}_1$ dressing it with an arbitrary elliptic function in $q$ a huge class
of new eigenfunctions can be obtained. This arbitrariness is lifted by the demand that the 
eigenfunction we are after diagonalize both operators ${\frak S}_{(0,s)}$ and ${\frak S}_{(s,0)}$
and in particular are symmetric with respect to exchanging $p$ and $q$.

The RS models have a long history of rich connections with gauge theories in various dimensions~(see {\it e.g.}~\cite{gorsky,Gorsky:1993pe}).
Nevertheless, for general values of $(p, q, t)$  determining the exact eigenfunctions and eigenvalues of the difference operators is still an open problem.
For some natural  limits of the parameters the eigenfunctions
 are well known. Curiously, 
many of the same limits have independent physical interest, because they lead
to a supersymmetry enhancement of the $\S^3 \times \S^1$ partition function.
One can systematically classify the limits of the index that enjoy enhanced supersymmetry,
and relate them to integrable models. We will shortly review some of the salient results in this direction.
Physical properties of theories of class ${\cal S}$ impose additional constraints on $\phi_\lambda ({\bf z})$. For example, since some of the theories have known Lagrangian descriprion the indices can be explicitly computed as integrals of elliptic Gamma functions and the results have to match the expressions
evaluated using the eigenfunctions. Exploiting the known expressions for the eigenfunctions for specialized values of the parameters and the additional physical constraints one can set up 
a perturbative scheme around the known results to compute the eigenfunctions for general values of the parameters~\cite{Gaiotto:2012xa,Razamat:2013qfa}.

\

\noindent We now turn to discuss several useful limits of the index for which explicit expressions 
for eigenfunctions are known.

\

\noindent {\bf Schur index}

\

\noindent 
The trace formula (\ref{inddef}) that defines the general index can be written in the following equivalent form 
(we suppress flavor fugacities to avoid cluttering):
\begin{equation}\label{inddef2}
{\cal I} (q, p, t)=\mathrm{Tr} (-1)^F\,
 p^{  \half \delta^{1 }_- }\,
 q^{ \half \delta^{1}_+ }\,
 t^{R+r}\,
   e^{-\beta' \delta_{2 \dot -}}\, ,
   \end{equation}
where 
\bea \label{deltas}
2 \delta^1_{ +} & \colonequals   \{ {\cal Q}^1_ +  \, ,   ({\cal Q}^1_{+})^\dagger \}   = & E+ 2j_1-2 R-r  \geqslant 0 \\
 2 \delta^1_{ -}  &\colonequals \{ {\cal Q}^1_{ -}\, ,   ({\cal Q}^1_{ -})^\dagger \}   = & E- 2j_1-2 R-r \geqslant 0 \nonumber \\
 2 \delta_{2 \dot -} & \colonequals \{{ \widetilde {\cal Q}}_{2 \dot -}\, , ({\widetilde  {\cal Q}}_{2 \dot  -})^\dagger \}  = &E- 2j_2-2 R+r  \geqslant 0\, . \nonumber
\eea
The inequalities follow from unitarity of the representation and will be useful momentarily.
The equivalence of (\ref{inddef}) and (\ref{inddef2}) follows immediately by recalling that  only states
with $\delta_{2 \dot -} = 0$ contribute to the trace.
The Schur index is the ``unrefined'' index obtained by setting $q=t$.  \,.
One readily observes that on this slice the combination of conserved charges appearing in the trace formula commute with a second supercharge, 
${\cal Q}^1_-$,
in addition to the supercharge $\widetilde {\cal Q}_{2\dot{-}}$ that leaves invariant the general index.
As the $p$ dependence is  ${\cal Q}^1_-$-exact, it drops out, and we are left with
a simple expression that depends on $q$ alone,\footnote{In principle the Schur index might make sense also for {\it non-conformal} ${\cal N}=2$ theories quantized on $\S^3 \times \mathbb{R}$,
although we are not aware of a detailed analysis of the requisite deformations needed to define an ${\cal N}=2$ theory on such a curved background (the analysis of~\cite{Klare:2013dka} might be of help here).
The ${\cal N}=1$ analysis of \cite{Festuccia:2011ws}
 is not  sufficient, because the Schur index cannot be understood
as a special case of the ${\cal N}=1$ index.
Of course, in the non-conformal case one cannot relate $\S^3 \times \mathbb{R}$ to  $\mathbb{R}^4$ by a Weyl rescaling and there
is no state/operator map.
 }
\be \label{schurindex}
{\cal I}_{\rm Schur} \colonequals \mathrm{Tr} (-1)^F\, q^{E-R} \,.
\ee
The index counts operators with $\delta^1_ - = \delta_{2 \dot -} = 0$, or equivalently 
\be\label{schurconditions}
\widehat L_0 \colonequals \frac{E - (j_1 + j_2)}{2} - R = 0 \, ,\qquad {\cal Z} \colonequals j_1 - j_2 + r = 0 \,.
\ee
In fact, the unitarity inequalities in (\ref{deltas}) give $\widehat L_0 \geqslant \frac{ | {\cal Z} | }{2}$, 
so the first condition implies the second. We refer to operators obeying $\widehat L_0 = 0$ as
{\it Schur operators}.    A Schur operator  is annihilated by two Poincar\'e supercharges of \emph{opposite} chiralities (${\cal Q}_-^1$ and  $\widetilde {\cal Q}_{2 \dot -}$
in our conventions). This is a consistent condition because the supercharges have the same $SU(2)_R$ weight, and thus anticommute with each other. No analogous BPS condition exists in an $\NN=1$ supersymmetric theory, because the anticommutator of opposite-chirality supercharges necessarily yields a momentum operator, which annihilates only the identity.

\renewcommand{\arraystretch}{1.5}
\begin{table}
\centering
\begin{tabular}{|l|l|l|l|l|}
\hline \hline
Multiplet  & $\OO_{\rm Schur}$  & $h \colonequals \frac{E + (j_1 + j_2)}{2}$ & $r$  & Lagrangian  ``letters''  \\ 
\hline 
$\hat \BB_R$  &  $\Psi^{11\dots 1}$   &    $R$ &  $0$ & $Q$, $\tilde Q$ \\ 
\hline
$\DD_{R (0, j_2 )}$  &    $ \widetilde {\QQ}^1_{\dot +} \Psi^{11\dots 1}_{\dot  + \dots \dot  + }$ &   $R+ j_2 +1$  & $j_2 + \frac{1}{2}$  &  $Q$, $\tilde Q$, $\tilde \lambda^1_{\dot +}$ \\
\hline
$\bar \DD_{R (j_1, 0 )}$  & $ {\QQ}^1_{ +} \Psi^{11\dots 1}_{+   \dots +}$ &     $R+ j_1 +1$  & $-j_1 - \frac{1}{2}$  & $Q$, $\tilde Q$,  $\lambda^1_{+}$ \\
\hline
$\hat \CC_{R (j_1, j_2) }$ &   ${\QQ}^1_{+} \widetilde  {\QQ}^1_{\dot +} \Psi^{11\dots 1}_{+   \dots + \, \dot  + \dots \dot  + }$&   
 $R+ j_1 + j_2 +2$  & $j_2 - j_1$  &
 $D_{+ \dot +}^n Q$,  $D_{+ \dot +}^n \tilde Q$,  $D_{+ \dot +}^n \lambda^1_{+}$,
  $D_{+ \dot +}^n  \tilde \lambda^1_{\dot +}$ \\
\hline
\end{tabular}
\caption{\label{schurTable} This table summarizes the manner in which Schur operators fit into short multiplets of the $\NN=2$ superconformal algebra. 
We use the naming conventions for supermultiplets of Dolan and Osborn \cite{Dolan:2002zh}.
For each supermultiplet, we denote by $\Psi$ the superconformal primary. There is then a single conformal primary Schur operator ${\OO}_{\rm Schur}$, which in general is obtained by the action of some Poincar\'e supercharges on $\Psi$. We list the holomorphic dimension $h$ and $U(1)_r$ charge $r$ of ${\OO}_{\rm Schur}$ in terms of the quantum numbers $(R,j_1,j_2)$ that label the shortened multiplet (left-most column). We also indicate the schematic form that ${\OO}_{\rm Schur}$ can take in a Lagrangian theory by enumerating the elementary ``letters'' from which the operator may be built. We denote by $Q$ and $\tilde Q$ the complex scalar fields of a hypermultiplet, by $\lambda_{\alpha}^\II$  and $\tilde \lambda_{\dot \alpha}^\II$ the left- and right-moving fermions of a vector multiplet, and by $D_{\alpha \dot \alpha}$ the gauge-covariant derivatives. Note that while in a Lagrangian theory Schur operators
are  built from these letters, the converse is false -- not {\it all} gauge-invariant words of this kind are Schur operators, only
the special combinations with vanishing anomalous dimensions.  
} 
\end{table}

A summary of the different classes of Schur operators, organized according to  how they fit in shortened multiplets of the superconformal
algebra, is given in Table \ref{schurTable}~\cite{Beem:2013sza}.
The first line lists the half-BPS operators belonging to the Higgs branch ${\cal N}=1$ chiral ring,
which have $E =  2 R$ and $j_1 = j_2 = 0$.
In a Lagrangian theory, these are  operators of the schematic form $Q Q \dots \tilde Q \tilde Q$. 
The
$SU(2)_R$ highest weight component of the moment map operator $\mu^{11}$, which has $E = 2 R = 2$ 
(and transforms in the adjoint representation of the flavor group)
is in this class.
The second and third lines of the table list more general 
${\cal N}=1$ antichiral (respectively chiral) operators.
In a Lagrangian theory they may be obtained  by considering gauge-invariant words that contain   $\tilde \lambda^1_{\dot +}$ (respectively $\lambda^1_+$)
in addition to $Q$ and $\tilde Q$. Finally the forth line lists the most general class of Schur operators, belonging to supermultiplet obeying less familiar semishortening conditions. An important operator in this class is the  Noether current for the $SU(2)_R$ R-symmetry, which belongs to the same superconformal multiplet as the stress-energy tensor and is   universally present in any ${\cal N}=2$ SCF.  Its
 $J^{11} _{+ \dot +}$ component, with $E= 3$, $R=1$, $j_1  = j_2 = \half$, is a Schur operator.
 Finally, note that the half-BPS operators of the {\it Coulomb} branch chiral ring (of the form ${\rm Tr}\, \phi^k $
in a Lagrangian theory) are {\it not} Schur operators.

The Schur index earns its name from the fact that the wavefunctions are proportional to Schur polynomials,
and simple closed form expressions are available for all the ingredients that enter the TQFT formula  for the index (\ref{finindex}).
We will quote the full expressions below in the more general Macdonald limit.
The structure constants $C_\lambda (q) $ turn out to be inversely proportional to the {\it quantum dimension} of the representation $\lambda$.
One recognizes~\cite{Gadde:2011ik} the  TQFT of the index as the zero-area limit\footnote{On a surface of finite (non-zero) area, $q$-YM is not topological, but
it still admits a natural class ${\cal S}$ interpretation~\cite{Tachikawa:2012wi}  as the supersymmetric partition function
of the $(2, 0)$ theory on $\S^3\times \S^1\times {\cal C}$ where 
the UV curve ${\cal C}$ is kept of finite area~\cite{GMT}.}
of $q$-deformed $2d$ Yang-Mills theory~\cite{Aganagic:2004js}, which
can also be understood as an analytic continuation of Chern-Simons theory on ${\cal C} \times \S^1$. 
This observation has been reproduced by a top-down approach ~\cite{Kawano:2012up,Fukuda:2012jr},
starting from the $(2, 0)$ theory on the geometry $\S^3\times \S^1\times {\cal C}$, first reducing 
on $\S^1$ to obtain $5d$ YM, and then reducing further on $\S^3$ and using supersymmetric localization   to obtain a bosonic gauge theory on ${\cal C}$, which is argued to coincide with $q$-YM. 

In $q$-YM theory,  introducing flavor punctures correspond to 
fixing the holonomies of the gauge fields around the punctures. One 
can also define additional local operators by fixing the {\it dual} variables at the
 punctures~\cite{Witten:1991we} -- in the language of Chern-Simons theory on ${\cal C} \times \S^1$,
this corresponds to adding a Wilson loop along the temporal $\S^1$.
These operators are the natural  candidates to correspond to the surface defects discussed in the previous section~\cite{Gaiotto:2012xa, Alday:2013kda}.

Perhaps the most interesting fact about the Schur index
is that it can be viewed as the character of a $2d$ chiral algebra
canonically associated to the $4d$ SCFT~\cite{Beem:2013sza}, as we shall review in section \ref{chiralalg}. A related point is that the Schur index enjoys 
intriguing modular properties encoding conformal anomalies~\cite{Razamat:2012uv}. For example
the indices of a hypermultiplet and the vecor multiplets in the Schur limit become 
combinations of  theta functions,
\be
{\cal I}_H=\frac1{\theta(q^{\frac12}a;q)}\,,\qquad
\Delta_{Haar}({\bf z})\,{\cal I}_V({\bf z})= \frac1{n!} (q;q)^{2n-2} \prod_{i\neq j}\theta(qz_i/z_j;q)\,,
\ee  which have simple modular properties under
\be
q=e^{2\pi i\tau}\quad\to \quad q'=e^{-\frac{2\pi i}\tau},\qquad
z=e^{2\pi i \zeta}\quad\to\quad z'=e^{\frac{2\pi i \zeta}\tau}\,. 
\ee Here $\Delta_{Haar}$ is the Haar measure and we specialized
for concreteness to $SU(n)$ vector field.
 An index of the gauge theory is given by contour integrals with the integrand built from 
products of theta functions. The combination of theta functions in in the integrand, ${\cal I}_{integ.}({\bf z};q)$, always 
forms an elliptic function in the fugacities, ${\bf z}$ corresponding to the gauged symmetries,
\be
{\cal I}_{integ.}(q\,{\bf z};q) = {\cal I}_{integ.}({\bf z};q)\,.
\ee
The gauge fugacities ${\bf z}$ can be thus thought as taking values on a torus with modular 
parameter $\tau$.
The contour integral defining the Schur index of a gauge theory then can be thought of as an integral 
over a cycle of the torus while the index after modular transformation is given as an integral over the dual cycle. 
\begin{figure}[ht]
\begin{center}
\includegraphics[scale=.41]{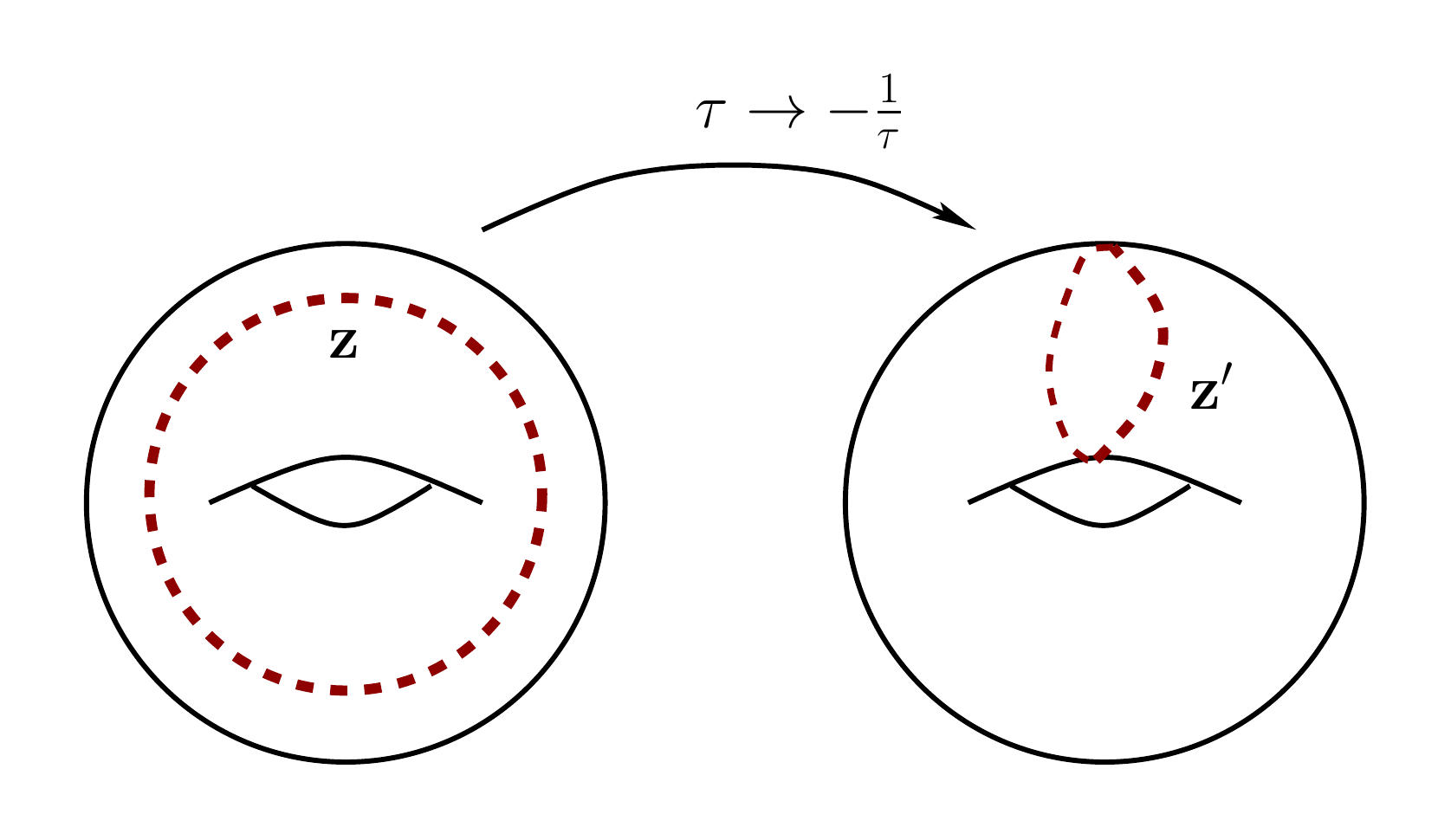}
\end{center}
\caption{The Schur index of a guage theory is given by an integral over fugacities ${\bf z}$ taking value in a torus with modular parameter $\tau$. After modular transformation, $\tau \to -\frac1\tau$, the index
is written as an integral over the dual cycle.}
\end{figure} These properties beg the question of the relation of the Schur index to {\it mock modular forms}, a relation which is yet to be explored.

\

\noindent {\bf Macdonald limit}

\

\noindent Taking $p\to 0$ in (\ref{inddef2}) is a well-defined limit,
thanks to positive-definiteness of the associated charge $\delta^1_-$.
The trace formula reads
\be
{\cal I}_{\rm Mac} (q, t) \colonequals {\rm Tr}_{\rm M} (-1)^F  q^{E-2R-r} t^{R+r}  =  {\rm Tr}_{\rm M} (-1)^F  q^{2 j_1} t^{R+r}   \, ,
\ee
where the subscript in the trace indicates that we are restricting by hand to the states with $\delta^1_- = 0$.
Clearly, we are concentrating on the operators that are also  annihilated  by the  supercharge  ${\cal Q}^1_{-}$, in addition to $\widetilde {\cal Q}_{2 \dot -}$.
  These are of course the same as the Schur operators, but  we are now refining their counting
  by  keeping track of the quantum number $R + r$. For $q=t$, we recover the Schur index.

This limit is mathematically very interesting. Our difference operators and our integration measure
become identical (up to conjugation)  to the well-studied  Macdonald difference operators and Macdonald measure \cite{Mac}.
The diagonalization problem is completely solved in terms of Macdonald polynomials, a beautiful two-parameter
generalization of the Schur polynomials. In the Macdonald limit we set the elliptic curve of the 
Ruijsenaars-Schneider model, $p$, to zero the integrable model becomes thus trigonometric (but 
still relativistic). For example, in the $A_1$ case after conjugation~\eqref{conj} the basic hamiltonian becomes,\footnote{Note that this is the same operator that we obtained in a quite different context of the reduction of the elliptic difference operator ${\frak S}_{(0,1)}$ to three dimensions~\ref{3dop}.} 
\be
{\cal H}_1 \cdot f(z) \sim \frac{1-t\,z^2}{1-z^2}\,f(q^{\frac12}z)+\frac{1-t\,z^{-2}}{1-z^{-2}}\, f(q^{-\frac12}\,z)\,.
\ee

We are then able to find
closed form expressions for the general wavefunctions and for  the structure constants \cite{Gadde:2011uv}. 
The wavefunction for a general choice of puncture (embedding) and representation now takes the following form,
\bea
\psi_{\Rf}^{\Lambda}({\bf z}_\Lambda ;q,t) = K_{\Lambda}({\bf z}_{\Lambda}; q,t)\,P_{\Rf}^\gf ( \mathrm{fug}_{\Lambda}(\mathbf{z}_{\Lambda}; t);q,t )~.
\eea
Here $P_{\Rf}^\gf (\mathbf{z}; q,t )$ are the Macdonald polynomials labeled by finite dimensional representatioins $\Rf$ of Lie algebra $\gf$ and orthonormal under the Macdonald measure, which, {\it e.g.}, for $\gf={\frak {su}}(n)$ is given by,
\be\label{mm}
\Delta_{q,t}({\bf z})=\frac{1}{n!}\prod_{i\neq j}\frac{(z_i/z_j;q)}{(tz_i/z_j;q)}\,.
\ee
The \emph{$K$-factors} admit a compact expression as a plethystic exponential \cite{Mekareeya:2012tn},
\be
K_{\Lambda}(\mathbf{z}_{\Lambda};q,t) = \PE \left[ \sum_j \frac{t^{j+1}}{1-q} \goodchi^{\hhf_\Lambda} _{\RR_j^{(\mathrm{adj})}}(\mathbf{z}_{\Lambda}) \right]~,
\ee
where the summation is over the terms appearing in the decomposition of Eqn. \eqref{generaldecomposition} applied to the adjoint representation,
\begin{equation}
\label{adjdecomposition2}
\mathrm{adj}_{\gf} = \bigoplus_j \RR_j^{(\mathrm{adj})} \otimes V_j~.
\end{equation}
$\goodchi^{\hhf_\Lambda} _{\RR_j^{(\mathrm{adj})}}({\mathbf z})$ is the Schur polynomial of Lie algebra $\hhf_\Lambda$ corresponding to representation $\RR_j^{(\mathrm{adj})}$.
 For the maximal puncture, corresponding to the trivial embedding $\Lambda_{\rm max} \equiv 0$, the wavefunction reads,
\begin{equation} 
\label{eq:psi_max}
\psi_{\Rf}^{ \Lambda_{\rm max} }({\bf x} ;q,t) =K_{\rm max} ({\bf x}; q,t) \, P_\Rf^\gf ( {\bf x} ;q,t) \,, \quad K_{\rm max} ({\bf x}; q,t) \ceq \PE \left[ \frac{t}{1-q} \chi_{\rm adj} ^\gf ( {\bf x} ) \right]~.
\end{equation}
At the other extreme, for the principal embedding $\Lambda = \rho$, the decomposition of Eqn. \eqref{adjdecomposition2} reads
\be
\label{adjoint_decomposition_principal}
\mathrm{adj}_{\gf} = \bigoplus_{i=1}^{{\rm rank} \, \gf} V_{d_i -1}~,
\ee
where $\{ d_i \}$ are the degrees of invariants of $\gf$, so in particular $d_i = i +1$ for $\suf(n)$. We then find
\be
\label{eq:psi_rho}
\psi^\rho_\Rf (q,t) = \PE \left[ \sum_{i}^{{\rm rank} \, \gf} \frac{t^{d_i} }{1-q} \right] P^\gf_\Rf ( \mathrm{fug}_{\rho}(t) )~.
\ee
For $\gf = \suf(n)$, the fugacity assignment associated to the principal embedding takes a particularly simple form,
\begin{equation}
\label{eq:principle_fugacities}
\mathrm{fug}_{\rho}(t) = (t^{\frac{n-1}{2} }, t^{\frac{n-3}{2}}, \dots t^{-\frac{n-1}{2} })~.
\end{equation}
Provided that,
\be
C_\Rf(q)^{-1} =  \psi^{\rho}_\Rf (q,t)\,,
\ee we thus obtain an expression for the Macdonald index of any class ${\cal S}$ theory with regular punctures,
\be
\label{MacUVcurve}
\II_{\rm Mac}(q,t; {\bf x} )= \sum_{\Rf} C_\Rf(q,t)^{2g-2+s}\prod_{i=1}^s \psi_{\Rf}^{\Lambda_i}({\bf x}_{\Lambda_i} ;q,t)~,
\ee with all the ingredients explicitly given above.

In the Macdonald limit, 
the TQFT of the index is recognized as a certain deformation of $q$-YM,
closely related to the {\it refined} Chern-Simons theory on ${\cal C} \times \S^1$ discussed
in~\cite{Aganagic:2011sg};  the refinement amounts to changing the measure in the path integral
of $q$-YM from Haar to Macdonald.

\

\noindent {\bf Hall-Littlewood limit}

\

\noindent  Proceeding one step further, we can 
  take the $q \to 0$ limit in the Macdonald index. The trace formula reads
 \be
 {\cal I}_{\rm HL} (t) \colonequals  {\rm Tr}_{\rm HL} (-1)^F \, t^{R+r}\, ,
 \ee
 where we are restricting the trace to states with $\delta^1_+ = \delta^1_- = 0$.
 In the  $q \to 0$ limit, Macdonald polynomials reduce to the much more manageable 
 Hall-Littlewood (HL) polynomials.    The HL index of theories of class ${\cal S}$ takes a relatively simple form: it is always a rational function of $t$.
 
 The HL index receives contributions from  operators annihilated by the three supercharges ${\cal Q}^1_{+}$, ${\cal Q}^1_{+}$ and ${\cal Q}_{2 \dot -}$.
This is  precisely the subset of Schur operators with $j_1 = 0$, corresponding to the $\hat {\cal B}$ and ${\cal D}$ multiplets,
listed in the first two rows of Table \ref{schurTable}.
Since such {\it Hall-Littlewood operators} are killed by both spinorial components of  ${\cal Q}^1_\alpha$, they are chiral\footnote{To be pedantic, {\it anti\/}chiral.}
with respect to an ${\cal N}=1$ subalgebra, and thus form a ring, which is consistent truncation of the full ${\cal N}=1$ chiral ring.
In a Lagrangian theory, they are composite  operators made with the complex hypermultiplet scalars $Q$ and $\tilde Q$ and the $\lambda^1_{\dot +}$
component of the gaugino, but  no derivatives.
 There is a further consistent truncation of the ring to operators with $j_2 =0$:
 this is the Higgs branch chiral ring, spanned by the bottom component of the  ${\hat {\cal  B}}_R$  multiplets.

For an ${\cal N}=2$ SCFTs associated to a linear quiver, one can show that
only the ${\hat {\cal B}}_R$ multiplets 
 contribute to the HL index.  This is the case because 
 the gauginos are in one-to-one correspondence
with the F-term constraints on the Higgs branch chiral operators, so their contribution
to the index (which comes with a minus sign) is precisely such to enforce those constraints.
 It follows that for linear quivers the HL index
coincides~\cite{Gadde:2011uv} 
with the Hilbert series of the Higgs branch (see {\it e.g.}~\cite{Gray:2008yu,Hanany:2008kn}).
The equivalence between the HL index and the Higgs branch Hilbert series
appears also to hold for the $T_n$ building blocks (see \cite{Gadde:2011uv}),
  and so by the same reasoning it extends to all class ${\cal S}$ theories associated to curves of 
 of {\it genus zero}.
 One can then use the HL index to compute the Hilbert series of multi-instanton moduli spaces for
$E_n$ groups~\cite{Hanany:2012dm,Gaiotto:2012uq}, which are quite intricate to compute using
 other methods (see {\it e.g.}~\cite{Keller:2012da}).   
 The HL index and the Higgs Hilbert series are not 
 the same for theories with genus one or higher, 
 where  ${\cal D}$ multiplets play a role.\footnote{
Assuming that the Higgs branch of the $4d$ theory of class ${\cal S}$ is isomorphic to the Higgs branch of the dimensionally reduced theory, we can consider the Coulomb index~\cite{Cremonesi:2014kwa,Cremonesi:2014vla,Razamat:2014pta} of the mirror dual theory (see section~\ref{threesec}). The $3d$ Coulomb index of the mirror coincides with the Hilbert series of the Higgs branch of theories of class ${\cal S}$ for any genus. We refer the reader to~\cite{Razamat:2014pta}
for further discussion of this issue.}

\

\noindent {\bf Coulomb limit}

\

\noindent There is another limit of the index that leads to supersymmetry enhancement: one takes
$t,\, p\to 0$ while keeping $q$ and 
$\frac{p\,q}{t}$ fixed. It is called the Coulomb limit
because in a Lagrangian theory the hypermultiplet single-particle index (\ref{spindices}) goes to zero;
the only supermultiplets that contribute in this limit are the short multiplets
 of type  $\bar {\cal E}_{-\ell(0,0)}$ (in the notations of \cite{Dolan:2002zh}), 
whose lowest components are the operators of the Coulomb branch chiral ring,
of the form ${\rm Tr  }\, \phi^k$. That there should exist a limit
of the general index for which only  $\{ \bar {\cal E}_{-\ell(0,0)} \}$ contribute
is {\it a priori} clear from the fact that these multiplets
do not appear in any of the recombination rules, so their multiplicities define an index.

In a Lagrangian theory with simple gauge group $G$, the
 Coulomb index is given by~\cite{Gadde:2011uv}
\be\label{maccent}
{\cal I}_C=\oint [d{\bf z}]_G\, \Delta_{q,\frac{pq}{t}}({\bf z})={\rm PE} \left[
\sum_{\ell\in\exp(G)}\widetilde {\cal I}_{\ell+1}\right]\,,
\ee where $\exp(G)$ stands for the set of exponents of $G$,
 $\Delta_{q,\frac{pq}{t}}({\bf z})$ is the Macdonald measure~\eqref{mm}
 (which arises by taking the Coulomb limit of the usual propagator measure),
and $\widetilde {\cal I}_{\ell+1}$ is the index of an individual $\bar {\cal E}_{-\ell(0,0)}$ multiplet.
This is a well-known mathematical equality, going by the name of the the Macdonald central term identity.
It can be understood physically as the statement that the Coulomb chiral ring
is freely generated by a set of operators in one-to-one correspondence
with the Casimir invariants of $G$, for example  $\{ {\rm Tr  }\, \phi^k \, \}$,  $k=2, \dots n$ for $G=SU(n)$.\footnote{The fact that the Coulomb branch is freely generated is known to be true by
inspection for theories of class ${\cal S}$ of type $A$ we discuss here, but is not obvious for
theories of type $D$ and $E$: it would be interesting to clarify this issue. We thank Y.~Tachikawa 
for this comment.}

\


\section{Some generalizations}\label{S:generalizations}

The discussion in previous sections can be extended and generalized in several ways. 
We will discuss some of the open problems in section \ref{conclusions}, while here let us briefly mention 
some of the work that has already appeared in the literature.

\begin{itemize}

\item In this review we have concentrated on class ${\cal S}$ theories of type $A$.
A similar analysis can be performed for theories of type $D$ and $E$.
Following our TQFT intuition the indices should be expressible in terms of 
a complete set of functions. The integrable models we discussed here 
for which the relevant set of functions for the $A$ case
is a set of eigenfunctions have natural generalizations to the $D$ and $E$ cases.
In particular the eigenfunctions for $D$ and $E$ cases are known in the Macdonald limit.
These eigenfunctions have been used to compute indices for the  three-punctured spheres $D$ type class ${\cal S}$ theories~\cite{Lemos:2012ph,Chacaltana:2013oka} and for the $E$ type
class ${\cal S}$ theories~\cite{Chacaltana:2014jba}.  One can also consider indices with outer-automorphism twists around the temporal $\S^1$ as was done in~\cite{Mekareeya:2012tn}.

\item Performing a different twist of the $6d$ theory while puttng it on a Riemann surface can result 
in a $4d$ theory with ${\cal N}=1$ supersymmetry rather than ${\cal N}=2$~\cite{Bah:2012dg}.
The resulting ${\cal N}=1$ theories are closely related to the ${\cal N}=2$ class theories and in particular 
their indices can be exactly computed resulting in expressions which are very similar to the ones discussed here~\cite{Beem:2012yn,Gadde:2013fma}. The ${\cal N}=1$ theories can be also built using outer-automorphism twists and the corresponding indices can be computed as was done in~\cite{Agarwal:2013uga}.

\item In the process of detemining the index we have found it useful to consider indices of theories
with surface defects. The theories of interest admit a variety of other supersymmetric defects in presence of which the index can be computed. For example, one can compute the Schur index in presence 
of supersymmetri line operator wrapping the $\S^1$~\cite{Dimofte:2011py,Gang:2012yr}.   Here the
answers are easily obtained in case of Wilson lines but in case of 't Hooft lines the computation
is much more involved~\cite{Gang:2012yr} if one chooses to perform the computation without making
use of S-duality. Other examples of extended objects involve domain walls~\cite{Gang:2012ff} and
more general surface defects than discussed here~\cite{Alday:2013kda,Bullimore:2014nla}.\footnote{
The index of theories of class ${\cal S}$ in presence of codimension two defects of the 6d theory wrapping the Riemann surface~\cite{Alday:2010vg}
has not been analyzed yet.
}

\item  Finally let us mention that the dualities satisfied by the theories of class ${\cal S}$ imply 
highly non-trivial identities satisfied by the superconformal indices. These identities take usually 
the form of equalities between different integrals of elliptic Gamma functions and or (infinite) sums 
of orthogonal functions. To give an example let us write down the index of the $SU(N)$ ${\cal N}=2$ 
SYM with $2N$ flavors. This theory corresponds to a sphere with two maximal and two minimal punctures and its index is proportional  to~\cite{Gadde:2010te},

\be
\oint \prod_{\ell=1}^{N-1}\frac{dz_\ell}{2\pi i z_\ell}\prod_{i\neq j}
\frac{\Gamma(\frac{pq}t z_i/z_j;p,q)}{
\Gamma(z_i/z_j;p,q)}\, \prod_{i=1}^N\prod_{\alpha,\beta=1}^{2N}
\Gamma(t^{\frac12} (z_i y_\alpha a)^{\pm1};p,q)
\Gamma(t^{\frac12}(z_i^{-1}  x_\beta b)^{\pm1};p,q)\,.
\ee Here $a$ and $b$ are fugacities for the $U(1)$ symmetries associated with the minimal punctures
and ${\bf x}$ with ${\bf y}$ are fugacities associated with the maximal punctures.
The S-duality exchanging the two minimal punctures implies that the above integral is invariant under exchange of $a$ and $b$. Mathematically this property is not at all obvious and was proven for the $SU(2)$
case in~\cite{debult}. As far as we know no mathematical proof for higher  rank cases exists as of this moment. Another simple example of an unproven identity following from S-duality propertied of the index
is the equality of the indices of $SO(2n+1)$ and $SP(n)$ ${\cal N}=4$ theories~\cite{Gadde:2009kb}. 

\end{itemize}


\section{Chiral algebras and the Schur index}\label{chiralalg}

In this section, we give a  brief outline of the structure discovered in \cite{Beem:2013sza}. 
 The basic claim is that any ${\cal N}=2$ SCFT admits a closed sector of operators and observables,
 isomorphic to a two-dimensional chiral algebra. The Schur index 
 is recognized as the character of this chiral algebra,
 \be \label{charschur}
 {\rm Tr}_{2d} \, (-1)^F q^{L_0}  \equiv {\cal I}_{\rm Schur} (q) \, .
 \ee
  To understand this surprising claim, we start with the following seemingly innocent observation.
The states that contribute to the Schur index  can be equivalently characterized as belonging
to the cohomology of  a single nilpotent supercharge, a linear combination of Poincar\'e and conformal supercharges,
\be
\qq \colonequals  \QQ^1_{-}+\tilde\SS^{\dot{-}2}\,.
\ee
Indeed, 
\be\label{defL0}
\{ \qq \, , \qq^\dagger \} = 2 \widehat L_0 = E-(j_1 + j_2) - 2 R  \,,
\ee
so the harmonic cohomology representatives  obey the Schur condition (\ref{schurconditions}). 
By the state/operator map, states are as always in correspondence with local operators inserted at the origin.
So Schur operators ${\cal O}_{\rm Schur}(0)$  inserted the origin  
belong to the cohomology of $\qq$.  

What is the cohomology
of $\qq$ more generally? One easily shows that ${\cal Z}$ defined in (\ref{schurconditions}) is $\qq$-exact,
so  a local operator can be $\qq$-closed only if it lies on the plane fixed by $j_1 - j_2$, which we
call the {\it chiral algebra plane}. We use  the complex coordinate $z$ (and its conjugate $\bar z$)
to parametrize the chiral algebra plane. The global conformal algebra on the chiral algebra plane
 is the standard
 $\slf(2) \times \ol {\slf(2)}$, with generators $L_n$ and $\bar L_n$, for $n= -1, 0, 1$,
 and is of course a subalgebra of the four-dimensional conformal algebra.
  For example,
 \be
 L_0 = \frac{E + j_1 + j_2}{2} \,, \quad \bar L_0 =  \frac{E - (j_1 + j_2)}{2}\,.
 \ee
 It turns out that 
\be
[ \qq\, , L_n]= 0 \, \quad {\rm but}   \, \quad [ \qq\, , \ol L_n] \neq 0 \, , 
\ee 
 so a Schur operator ${\cal O}_{\rm Schur}(z, \bar z)$
 inserted away from the origin is {\it not} $\qq$-closed.
 There is however a simple fix. We introduce
 a twisted algebra $\wh{\slf(2)}$  as the diagonal subalgebra of $\ol{\slf(2)}\times\suf(2)_R$,
\begin{equation}\label{eq:twisted_sl2_definition}
\Lh_{-1} \ceq \bar L_{-1}+\RR^-  \, , \qquad  \Lh_{0} \ceq \bar L_{0}-\RR  \, , \qquad \Lh_{+1} \ceq \bar L_{+1}-\RR^+  \, .
\end{equation}
(In retrospect, this explains  why the combination of charges in the first equation of \eqref{schurconditions} was denoted by $\Lh_0$).
Remarkably, the twisted generators $\Lh_{n}$ are $\qq$-exact. It follows that starting from a Schur operator
inserted at the origin, we can act with twisted translations to obtain a $\qq$-closed operator defined
at a generic point $(z, \bar z)$ on the chiral algebra plane,
\begin{equation}\label{eq:twisted_translation}
\OO(z,\zb)=e^{z L_{-1} + \zb \Lh_{-1} }\OO_{\rm Schur} (0,0)e^{-z L_{-1}- \zb \Lh_{-1}}~.
\end{equation}
A Schur operator is
necessarily an $\suf(2)_R$ highest weight state, carrying the maximum eigenvalue $R$ of the Cartan. Indeed, if this were not the case, states with greater values of $R$ would have negative $\Lh_0$ eigenvalue, violating unitarity. We denote the whole spin $k$ representation of $\suf(2)_R$ as $\OO^{(\II_1\cdots\II_{2k})}$, with $\II_i = 1,2$. Then the Schur operator is $\OO_{\rm Schur} = \OO^{11\cdots1}(0)$, and the twisted-translated operator at any other point is given by
\begin{equation}\label{displaced}
\OO(z, \zb) \colonequals u_{\II_1}(\zb)\,\cdots\,u_{\II_{2k} }(\zb) \; \OO^{(\II_1\cdots\II_{2k})} (z, \zb) \,, \qquad\quad u_{\II} (\zb) \colonequals (1, \zb) \,.
\end{equation}
By construction, such an operator is annihilated by $\qq$, and $\qq$-exactness of $\Lh_{-1}$ 
implies that its $\zb$ dependence is $\qq$-exact. It follows that the cohomology class of the twisted-translated operator defines a purely meromorphic operator,
\begin{equation}\label{eq:cohomology_to_chiral}
[\OO(z,\zb)]_\qq~~~\leadsto~~~\OO(z)~.
\end{equation}
Operators constructed in this manner have correlation functions that are meromorphic functions of the insertion points, and enjoy well-defined meromorphic OPEs at the level of the cohomology. These are precisely the ingredients that define a two-dimensional chiral algebra!
The relation (\ref{charschur}) of the chiral algebra character with  the Schur index follows at  once by observing that  
$\Lh_0 = 0$ implies $L_0 = E-R$, so  the trace formula (\ref{schurindex}) that defines the Schur index is reproduced.

\bigskip

There is a rich dictionary related properties of the $4d$ SCFT with properties of its associated chiral algebra.
Let us briefly mention some universal features of this correspondence:
\begin{itemize}
\item
The global $\mf{sl}(2)$ symmetry is enhanced to the full {Virasoro} symmetry,  with the $2d$ holomorphic stress tensor 
$T(z)$ arising from the Schur operator in the $SU(2)_R$ conserved current,
$
T(z)\colonequals [{\JJ}_R(z,\bar z)]_{\qq}\,.
$
The $2d$ central charge is given by 
 \be
  c_{2d} = - 12  \, c_{4d}  \, ,
 \ee
where $c_{4d}$ is  one of  conformal anomaly coefficients of the $4d$ theory (the one associated to the Weyl tensor squared).
\item The global flavor symmetry of the SCFT is enhanced to 
an affine symmetry in the associated chiral algebra, with the affine current 
 $J(z)$ arising from the moment map operator,
$
J(z)\colonequals [M(z,\bar z)]_{\qq} \,.
$
The $2d$ level is related to the $4d$ level by another universal relation,
\be
 k_{2d} = - \frac{k_{4d}}{2 }  \,.
\ee
\item The generators of the HL chiral ring give rise to generators of the chiral algebra.
Remarkably, the geometry of the $4d$ Higgs branch is encoded algebraically in 
vacuum module of the chiral algebra: 
  Higgs branch relations
correspond to null states.
\end{itemize}

Free SCFTs are associated to free chiral algebras. The free hypermultiplet corresponds
to the chiral algebra of symplectic bosons $(q, \tilde q)$, of weights $(\half,\half)$, while
the free vector multiplet  corresponds to a $(b, c)$ ghost system of weights $(1,0)$.

There is also a chiral algebra counterpart of the index gauging prescription  (\ref{generalgauging}).
We start with a SCFT ${\cal T}$, whose chiral algebra $\chi[{\cal T}]$ is known,
and define a new SCFT ${\cal T}_G$ by gauging a subgroup of the flavor symmetry, such that the gauge coupling
is exactly marginal. A naive guess for finding  the chiral algebra  associated ${\cal T}_G$ is to 
take the  tensor product of  $\chi[{\cal T}]$  with  a $(b^A c_A)$ ghost system
in the adjoint representation of $G$, and  restrict to gauge singlets. This would be the direct
analog of (\ref{generalgauging}), and is 
indeed the correct answer at zero gauge coupling. But at finite coupling, some of the Schur states are lifted
 and the chiral algebra must be smaller. There is an elegant prescription to find the quantum chiral algebra:
 one is instructed to  pass to the cohomology of 
\be
Q_{\rm BRST} \colonequals \oint \frac{dz}{2 \pi i} \, j_{\rm BRST} (z) \, , \quad  j_{\rm BRST} \colonequals c_A \left[ J^A - \frac{1}{2}  f^{AB}_{\quad C} \, c_B   b^C
\right ] \, ,
\ee
where $J^A$ is the $G$ affine current of $\chi[ {\cal T}]$.
This BRST operator is nilpotent  precisely when the $\beta_G = 0$, which amounts to $k_{2d} = - 2 h^\vee$, where $h^\vee$ is the dual Coxeter
number of $G$. By this prescription, we can in principle find $\chi[ {\cal T} ]$ for
 any  {Lagrangian} SCFT ${\cal T}$.
 
 The chiral algebra contains  much more information that the Schur index.
 The state space of the chiral algebra can be regarded as a ``categorification'' of the Schur index: 
 it consists of the cohomology classes of $\qq$, whereas the index only counts such cohomology classes with signs,
 and so 
 it knows about sets of short multiplets that are kinematically allowed to recombine but do not. In addition, there may be multiplets that cannot recombine but nonetheless make accidentally cancelling contributions to the index, and these are also seen in the categorification.
And of course, the chiral algebra structure goes well beyond categorification -- it is a rich algebraic system that also encodes the OPE coefficients of the Schur operators, and is subject to non-trivial associativity constraints.

For theories of class ${\cal S}$, there is a 
  generalized topological quantum field theory that associates to a decorated Riemann surface the corresponding
 chiral algebra. Associativity of the gluing of Riemann surfaces imposes highly non-trivial requirements on the chiral
 algebras of the elementary building blocks $T_n$. Finally, let us mention that the task of reducing
 the rank of a puncture can be accomplished directly in the chiral algebra setting, by a generalization of  
 quantum  Drinfeld-Sokolov reduction.
 We refer to  \cite{Beem:2013sza, classSchiral} for a detailed discussion 
 of this very rich structure.

\section{Some open questions}

\label{conclusions}

We conclude by discussing some open problems and 
possible generalizations of the  topics discussed in this review.
The main focus of this review was the partition function on $\S^3\times \S^1$ for ${\cal N}=2$ 
superconformal theories in four dimensions: generalizations and extensions of our logic 
can be entertained by relaxing each of these qualifiers. 

\begin{itemize}
\item {\bf More partition functions} 
A rather natural generalization is to consider indices with the theory quantized on more generic manifolds, 
{\it i.e.} ${\cal M}_3\times \S^1$. For example, one can take ${\cal M}_3=\S^3/{\mathbb Z}_r$, the lens space~\cite{Benini:2011nc}. The superconformal index discussed in this paper is a special case of the partition functions
defined using this sequence of manifolds, $r=1$. The lens space with $r>1$ has a non-contractable cycle and thus is sensitive to non-local objects in the theory. In particular, unlike the superconformal index
it can distinguish theories differing by choices of allowed line operators and/or by choices of the global
structure of the gauge groups.  One would expect that as long as the manifold
on which the  partition function is computed has an $\S^1$ the arguments of this paper
 can be reiterated. In particular the partition functions in these cases should be computable 
by a $2d$ TQFT. This has been discussed in the case of the lens space~\cite{Alday:2013rs,Razamat:2013jxa}, and it would be interesting 
to extend the analysis to other partition functions, {\it e.g.} ${\mathbb T}^2\times \S^2$~\cite{Closset:2013vra,Closset:2013sxa}.

\item {\bf More theories}

The superconformal index is not yet fixed for {\it all} ${\cal N}=2$ theories in $4d$. For example,
we do not know at the moment how to compute the index depending on the most general set of fugacities
for Argyres-Douglas theories and theories corresponding to Riemann surfaces with irregular singularieties~\cite{Xie:2012hs}.\footnote{See however~\cite{DelZotto:2014kka} for some recent discussion.}
It would be very interesting to fill this gap in our current understanding. To do so it might be useful
to exploit the chiral algebra associated to these theories and its relation to the (Schur) index.  
Another, related, question is what kind of partition functions can be exactly computed for ${\cal N}=2$ theories which are not superconformal.

\item {\bf Properties of the index}

The indices which we can compute have many interesting  properties not all of which were  sufficiently well studied.  For example, the $4d$ indices have factorization properties~\cite{Yoshida:2014qwa,Peelaers:2014ima} similar to the 
ones studied for the partition functions of $3d$ theories~\cite{Pasquetti:2011fj,Beem:2012mb,Cecotti:2013mba}. Another example is that of modular properties the indices have under non linear transformations of some of the chemical potentials~\cite{Spiridonov:2012ww,Razamat:2012uv} (see also~\cite{Aharony:2013dha,DiPietro:2014bca,Ardehali:2014esa}).

\item {\bf Less supersymmetry and/or other space-time dimensions}

A very important open question 
is whether the methodology which allowed us to fix the index of a large class of ${\cal N}=2$ theories
can be applied to theories with less supersymmetry and/or theories in different space-time dimensions:
this remains to be seen.

\item {\bf Relations to mathematics}

The superconformal index is directly related to different branches of exciting mathematics. To list 
just couple examples: it is a gold mine for extracting identities satisfied by elliptic hypergeometric integrals; and it is closely related to  quantum mechanical integrable systems with their very rich 
mathematical structure.  There is a real chance here for a mutually beneficial dialogue between the
 mathematics community working on these topics and the physics community.

\end{itemize}

\bigskip
\section*{Acknowledgments}

It is a great pleasure to thank 
Chris Beem,  Abhijit Gadde, Davide Gaiotto, Madalena Lemos, Pedro Liendo,  Wolfger Peelaers, Elli Pomoni,
Brian Willett, and Wenbin Yan, 
for very enjoyable collaboration and  countless 
 discussions on the material reviewed here.
 We  thank Davide Gaiotto and Yuji Tachikawa for useful comments on the draft.
LR thanks
 the Simons Foundation and the Solomon Guggenheim Foundation for their generous support. He is grateful to the IAS,
Princeton, and to the KITP, Santa Barbara,  for their wonderful hospitality during his sabbatical leave. LR is also supported
in part by the National Science Foundation under Grant No. NSF PHY1316617.
SSR gratefully acknowledges support from the Martin~A.~Chooljian and Helen Chooljian membership
at the Institute for Advanced Study. The research of SSR was also partially supported by
National Science Foundation under Grant No. PHY-0969448, and by  "Research in Theoretical High Energy Physics" grant DOE-SC00010008.  SSR  would like to thank KITP, Santa Barbara, and the Simons Center, Stony Brook, for hospitality and support during different stages of this work. 
\appendix
\section{Plethystics}\label{notationssec}

In this appendix we collect the definitions
of some special functions and combinatorial  objects used in the bulk of the review. 
The Pochammer symbol is defined as
\be
(z;\,p) \colonequals \prod_{\ell=0}^\infty (1-z\,p^{\ell})\,.
\ee The theta-function is given by
\be
\theta(z;\,p) \colonequals (z;\,p)\,(p\,z^{-1};\,p)\,.
\ee 
The plethystic exponential is given by
\be
{\rm PE} \left[f(x,y,\cdots)\right] \colonequals  \exp\left[\sum_{\ell=1}^\infty \frac{1}{\ell}f(x^\ell,y^\ell,\cdots)\right]\,.
\ee
In particular 
\be
{\rm PE}[x]=\frac{1}{1-x}\,,\qquad {\rm PE}[-x]=1-x\,.
\ee
The inverse of the plethystic exponential  is the plethystic logarythm, given by
\be
{\rm PL}\left[f(x,y,\cdots)\right] \colonequals \sum_{\ell=1}^\infty \frac{\mu(\ell)}{\ell}\,\ln f(x^\ell,y^\ell,\cdots)\,,
\ee  where $\mu(\ell)$ is the Mobius mu-function.
Finally the elliptic Gamma function is defined as
\be
\Gamma(z;\,p,\,q) \colonequals {\rm PE} \left[\frac{z-\frac{p\,q}{z}}{(1-p)(1-q)}\right]=\prod_{i,j=0}^\infty\frac{1-p^{i+1}q^{j+1}z^{-1}}{1-p^iq^j\,z}\,.
\ee

\section{${\cal N}=2$ superconformal representation theory}

\label{shorteningappendix}

In this appendix (adapted from~\cite{Beem:2013sza})
we review 
the classification of short representations of the four-dimensional $\NN=2$ superconformal algebra \cite{Dobrev:1985qv,Dolan:2002zh, Kinney:2005ej}.

Short representations occur when the norm of a superconformal descendant state in what would otherwise be a long representation is rendered null by a conspiracy of quantum numbers. The unitarity bounds for a superconformal primary operator are given by
\be\label{eq:unit_bounds}
\begin{alignedat}{3}
E	&\geqslant  E_i~,&\qquad				&&				 j_i&\neq0~,\\
E	&=	E_i-&2~~\mbox{~or~}~~E&\geqslant & E_i~,	\qquad j_i&=0~,\\
\end{alignedat}
\ee
where we have defined
\be
E_1=2+2j_1+2R+ r~, \qquad E_2=2+2j_2+2R- r~,
\ee
and short representations occur when one or more of these bounds are saturated. The different ways in which this can happen correspond to different combinations of Poincar\'e supercharges that will annihilate the superconformal primary state in the representation. There are two types of shortening conditions, each of which has four incarnations corresponding to an $SU(2)_R$ doublet's worth of conditions for each supercharge chirality:
\bea\label{eq:constitutent_shortening_conditions}
\BB^\II&:&\qquad \QQ^\II_{\alpha}|\psi\rangle=0~,\quad\alpha=1,2\\
{\bar \BB}_\II&:&\qquad \wt\QQ_{\II\dot\alpha}|\psi\rangle=0~,\quad\dot\alpha=1,2\\
\CC^\II&:&\qquad
\begin{cases} \epsilon^{\a\b}\QQ^\II_{\alpha}&|\psi\rangle_\beta=0~,\quad j_1\neq0\\
\epsilon^{\a\b}\QQ^\II_{\alpha}\QQ^\II_{\beta}&|\psi\rangle=0~,\quad j_1=0
\end{cases}~,\\
{\bar\CC}_\II&:&\qquad
\begin{cases} \epsilon^{\alpha\beta}\wt\QQ_{\II\alpha}&|\psi\rangle_\beta=0~,\quad j_2\neq0\\
\epsilon^{\alpha\beta}\wt\QQ_{\II\alpha}\wt\QQ_{\II\beta}&|\psi\rangle=0~,\quad j_2=0
\end{cases}~,
\eea
The different admissible combinations of shortening conditions that can be simultaneously realized by a single unitary representation are summarized in Table \ref{Tab:shortening}, where the reader can also find the precise relations that must be satisfied by the quantum numbers $(E,j_1,j_2,r,R)$ of the superconformal primary operator, as well as the notations used to designate the different representations in \cite{Dolan:2002zh} (DO) and \cite{Kinney:2005ej} (KMMR).\footnote{We follow the R-charge conventions of DO.}

\begin{table}[t]
\begin{centering}
\renewcommand{\arraystretch}{1.3}
\begin{tabular}{|l|l|l|l|}
\hline
Shortening & Quantum Number Relations & DO & KMMR \tabularnewline
\hline
\hline 
$\varnothing$	
					& \makebox[3.8cm][l]	{$E\geqslant {\rm max}(E_1,E_2)$}								  	& $\AA^\Delta_{R,r(j_1,j_2)}$ & ${\bf aa}_{\Delta,j_1,j_2,r,R}$ 	\tabularnewline
\hline 
$\BB^1$ 							& \makebox[3.8cm][l]{$E=2R+r$}			\makebox[3cm][l]{$j_1=0$}	  	& $\BB_{R,r(0,j_2)}$ 		& ${\bf ba}_{0,j_2,r,R}$ 		\tabularnewline
\hline 
$\bar\BB_2$							& \makebox[3.8cm][l]{$E=2R-r$}			\makebox[3cm][l]{$j_2=0$}   	& $\bar{\BB}_{R,r(j_1,0)}$ 	& ${\bf ab}_{j_1,0,r,R}$ 		\tabularnewline
\hline 
$\BB^1\cap\BB^2$  					& \makebox[3.8cm][l]{$E=r$}  			\makebox[3cm][l]{$R=0$}  		& $\EE_{r(0,j_2)}$ 			& ${\bf ba}_{0,j_2,r,0}$ 		\tabularnewline
\hline
$\bar\BB_1\cap\bar\BB_2$  			& \makebox[3.8cm][l]{$E=-r$}  			\makebox[3cm][l]{$R=0$}  		& $\bar \EE_{r(j_1,0)}$ 	& ${\bf ab}_{j_1,0,r,0}$ 		\tabularnewline
\hline 
$\BB^1\cap\bar\BB_{2}$  			& \makebox[3.8cm][l]{$E=2R$}  			\makebox[3cm][l]{$j_1=j_2=r=0$}	& $\hat{\BB}_{R}$ 			& ${\bf bb}_{0,0,0,R}$ 			\tabularnewline
\hline\hline 
$\CC^1$ 							& \makebox[3.8cm][l]{$E=2+2j_1+2R+r$}  									& $\CC_{R,r(j_1,j_2)}$ 		& ${\bf ca}_{j_1,j_2,r,R}$ 		\tabularnewline
\hline 
$\bar\CC_2$  						& \makebox[3.8cm][l]{$E=2+2 j_2+2R-r$}  								& $\bar\CC_{R,r(j_1,j_2)}$ 	& ${\bf ac}_{j_1,j_2,r,R}$		\tabularnewline
\hline
$\CC^1\cap\CC^2$  					& \makebox[3.8cm][l]{$E=2+2j_1+r$}  	\makebox[3cm][l]{$R=0$}  		& $\CC_{0,r(j_1,j_2)}$ 		& ${\bf ca}_{j_1,j_2,r,0}$ 		\tabularnewline
\hline 
$\bar\CC_1\cap\bar\CC_2$			& \makebox[3.8cm][l]{$E=2+2 j_2-r$} 	\makebox[3cm][l]{$R=0$}  	 	& $\bar\CC_{0,r(j_1,j_2)}$ 	& ${\bf ac}_{j_1,j_2,r,0}$ 		\tabularnewline
\hline 
$\CC^1\cap\bar\CC_2$  				& \makebox[3.8cm][l]{$E=2+2R+j_1+j_2$}	\makebox[3cm][l]{$r=j_2-j_1$}   & $\hat{\CC}_{R(j_1,j_2)}$ 	& ${\bf cc}_{j_1,j_2,j_2-j_1,R}$\tabularnewline
\hline\hline 
$\BB^1\cap\bar\CC_2$  				& \makebox[3.8cm][l]{$E=1+2R+j_2$}  	\makebox[3cm][l]{$r=j_2+1$}   	& $\DD_{R(0,j_2)}$ 			& ${\bf bc}_{0,j_2,j_2+1,R}$	\tabularnewline
\hline 
$\bar\BB_2\cap\CC^1$  				& \makebox[3.8cm][l]{$E=1+2R+j_1$}		\makebox[3cm][l]{$-r=j_1+1$}    & $\bar\DD_{R(j_1,0)}$ 		& ${\bf cb}_{j_1,0,-j_1-1,R}$ 	\tabularnewline
\hline 
$\BB^1\cap\BB^2\cap\bar\CC_2$  		& \makebox[3.8cm][l]{$E=r=1+j_2$} 		\makebox[2.5cm][l]{$r=j_2+1$} 	\makebox[1.5cm][l]{$R=0$}	& $\DD_{0(0,j_2)}$ 		& ${\bf bc}_{0,j_2,j_2+1,0}$  \tabularnewline
\hline
$\CC^1\cap\bar\BB_1\cap\bar\BB_2$  	& \makebox[3.8cm][l]{$E=-r=1+j_1$}  	\makebox[2.5cm][l]{$-r=j_1+1$} 	\makebox[1.5cm][l]{$R=0$}	& $\bar\DD_{0(j_1,0)}$ 	& ${\bf cb}_{j_1,0,-j_1-1,0}$ \tabularnewline
\hline
\end{tabular}
\par\end{centering}
\caption{\label{Tab:shortening}Unitary irreducible representations of the $\NN=2$ superconformal algebra.}
\end{table}

At the level of group theory, it is possible for a collection of short representations to recombine into a generic long representation whose dimension is equal to one of the unitarity bounds of \eqref{eq:unit_bounds}. In the DO notation, the generic recombinations are as follows:
\bea\label{eq:recombination}
\AA_{R,r(j_1,j_2)}^{2R+r+2+2j_1}&\simeq& \CC_{R,r(j_1,j_2)}\oplus \CC_{R+\hf,r+\hf(j_1-\hf,j_2)}~,\\
\AA_{R,r(j_1,j_2)}^{2R-r+2+2j_2}&\simeq& \bar\CC_{R,r(j_1,j_2)}\oplus \bar\CC_{R+\hf,r-\hf(j_1,j_2-\hf)}~,\\
\AA_{R,j_1-j_2(j_1,j_2)}^{2R+j_1+j_2+2}&\simeq& \hat\CC_{R(j_1,j_2)}\oplus \hat\CC_{R+\hf(j_1-\hf,j_2)}\oplus\hat\CC_{R+\hf(j_1,j_2-\hf)}\oplus\hat\CC_{R+1(j_1-\hf,j_2-\hf)}~.
\eea
There are special cases when the quantum numbers of the long multiplet at threshold are such that some Lorentz quantum numbers in \eqref{eq:recombination} would be negative and unphysical:
\bea\label{eq:special_recombination}
\AA_{R,r(0,j_2)}^{2R+r+2} 			& \simeq & 			\CC_{R,r(0,j_2)} 		\oplus 		\BB_{R+1,r+\hf(0,j_2)}~,\\
\AA_{R,r(j_1,0)}^{2R-r+2} 			& \simeq & 			\bar\CC_{R,r(j_1,0)} 	\oplus 		\bar\BB_{R+1,r-\hf(j_1,0)}~,\\
\AA_{R,-j_2(0,j_2)}^{2R+j_2+2} 			& \simeq & 			\hat\CC_{R(0,j_2)} 	\oplus 		\DD_{R+1(0,j_2)} \oplus 		\hat\CC_{R+\hf(0,j_2-\hf)} 		\oplus 		\DD_{R+\frac{3}{2}(0,j_2-\hf)}~,\\
\AA_{R,j_1(j_1,0)}^{2R+j_1+2} 			& \simeq & 			\hat\CC_{R(j_1,0)} 	\oplus 		\hat\CC_{R+\hf(j_1-\hf,0)} \oplus 		\bar\DD_{R+1(j_1,0)} 	\oplus 		\bar\DD_{R+\frac{3}{2}(j_1-\hf,0)}~,\\
\AA_{R,0(0,0)}^{2R+2} 			& \simeq & 			\hat\CC_{R(0,0)} 	\oplus 		\DD_{R+1(0,0)} 	\oplus 		\bar\DD_{R+1(0,0)} 	\oplus 		\hat\BB_{R+2}~.
\eea
The last three recombinations involve multiplets that make an appearance in the associated chiral algebra described in this work. Note that the $\EE$, $\bar\EE$, $\hat  {\cal B}_{\frac{1}{2}}$, 
$\hat  {\cal B}_{1}$, $\hat  {\cal B}_{\frac{3}{2}}$, $\DD_0$,  $\bar\DD_0$, $\DD_{\frac{1}{2}}$ and $\bar\DD_{\frac{1}{2}}$ multiplets can never recombine, along with $\BB_{\frac12,r(0,j_2)}$ and $\bar\BB_{\frac12,r(j_1,0)}$.

\bibliography{Review}
\bibliographystyle{JHEP}

\end{document}